\documentclass[twoside,twocolumn,9pt]{article}
\usepackage{extsizes}
\usepackage[super,sort&compress,comma]{natbib}
\usepackage[left=1.5cm, right=1.5cm, top=1.785cm, bottom=2.0cm]{geometry}
\usepackage{mathptmx}
\usepackage{sectsty}
\usepackage{graphicx} 
\usepackage{lastpage}
\usepackage[format=plain,justification=justified,singlelinecheck=false,font={stretch=1.125,small,sf},labelfont=bf,labelsep=space]{caption}
\usepackage{fancyhdr}
\usepackage{fnpos}
\usepackage[english]{babel}
\addto{\captionsenglish}{%
  
}
\usepackage{array}
\usepackage{droidsans}
\usepackage{charter}
\usepackage[T1]{fontenc}
\usepackage{subcaption}
\usepackage[usenames,dvipsnames]{xcolor}
\usepackage{setspace}
\usepackage[compact]{titlesec}
\usepackage{hyperref}

\usepackage{amsmath}
\usepackage{amssymb}
\usepackage{booktabs}

\definecolor{cream}{RGB}{222,217,201}

\begin{document}

\pagestyle{fancy}
\thispagestyle{plain}
\fancypagestyle{plain}{
\renewcommand{\headrulewidth}{0pt}
}

\makeFNbottom
\makeatletter
\renewcommand\LARGE{\@setfontsize\LARGE{15pt}{17}}
\renewcommand\Large{\@setfontsize\Large{12pt}{14}}
\renewcommand\large{\@setfontsize\large{10pt}{12}}
\renewcommand\footnotesize{\@setfontsize\footnotesize{7pt}{10}}
\makeatother

\renewcommand{\thefootnote}{\fnsymbol{footnote}}
\renewcommand\footnoterule{\vspace*{1pt}%
\color{cream}\hrule width 3.5in height 0.4pt \color{black}\vspace*{5pt}} 
\setcounter{secnumdepth}{3}

\makeatletter 
\renewcommand\@biblabel[1]{#1}            
\renewcommand\@makefntext[1]%
{\noindent\makebox[0pt][r]{\@thefnmark\,}#1}
\makeatother 
\renewcommand{\figurename}{\small{Fig.}~}
\sectionfont{\sffamily\Large}
\subsectionfont{\normalsize}
\subsubsectionfont{\bf}
\setstretch{1.125}
\setlength{\skip\footins}{0.8cm}
\setlength{\footnotesep}{0.25cm}
\setlength{\jot}{10pt}
\titlespacing*{\section}{0pt}{4pt}{4pt}
\titlespacing*{\subsection}{0pt}{15pt}{1pt}

\fancyfoot{}
\fancyfoot[LO,RE]{\vspace{-7.1pt}\includegraphics[height=9pt]{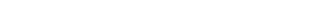}}
\fancyfoot[CO]{\vspace{-7.1pt}\hspace{13.2cm}\includegraphics{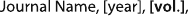}}
\fancyfoot[CE]{\vspace{-7.2pt}\hspace{-14.2cm}\includegraphics{head_foot/RF}}
\fancyfoot[RO]{\footnotesize{\sffamily{1--\pageref{LastPage} ~\textbar  \hspace{2pt}\thepage}}}
\fancyfoot[LE]{\footnotesize{\sffamily{\thepage~\textbar\hspace{3.45cm} 1--\pageref{LastPage}}}}
\fancyhead{}
\renewcommand{\headrulewidth}{0pt} 
\renewcommand{\footrulewidth}{0pt}
\setlength{\arrayrulewidth}{1pt}
\setlength{\columnsep}{6.5mm}
\setlength\bibsep{1pt}

\makeatletter 
\newlength{\figrulesep} 
\setlength{\figrulesep}{0.5\textfloatsep} 

\newcommand{\topfigrule}{\vspace*{-1pt}%
\noindent{\color{cream}\rule[-\figrulesep]{\columnwidth}{1.5pt}} }

\newcommand{\botfigrule}{\vspace*{-2pt}%
\noindent{\color{cream}\rule[\figrulesep]{\columnwidth}{1.5pt}} }

\newcommand{\dblfigrule}{\vspace*{-1pt}%
\noindent{\color{cream}\rule[-\figrulesep]{\textwidth}{1.5pt}} }

\makeatother

\twocolumn[
  \begin{@twocolumnfalse}
{\includegraphics[height=30pt]{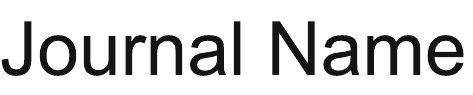}\hfill\raisebox{0pt}[0pt][0pt]{\includegraphics[height=55pt]{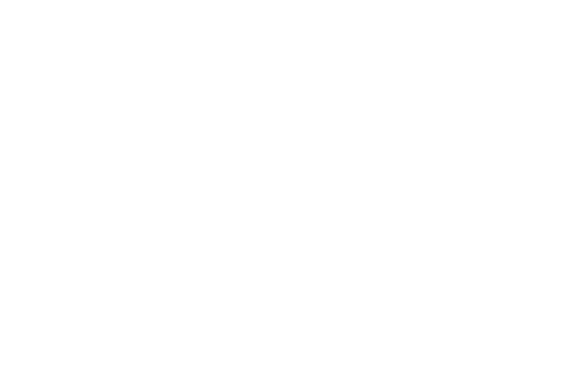}}\\[1ex]
\includegraphics[width=18.5cm]{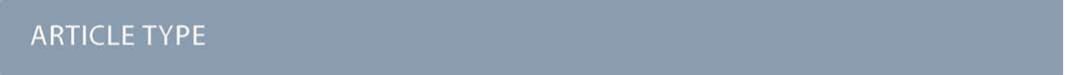}}\par
\vspace{1em}
\sffamily
\begin{tabular}{m{4.5cm} p{13.5cm} }

\includegraphics{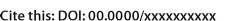} & \noindent\LARGE{\textbf{Equivariant Neural Prediction of the Stokes Resistance Tensors for Arbitrary Microparticle Shapes$^\dag$}} \\
\vspace{0.3cm} & \vspace{0.3cm} \\

 & \noindent\large{Sanjay Pradeep\textit{$^{a}$}, David Dandy\textit{$^{b}$}, Candace S. J. Tsai$^{\ast}$\textit{$^{c}$}, and Jeff D. Eldredge$^{\ast}$\textit{$^{a}$}} \\

\includegraphics{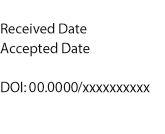} & \noindent\normalsize{%
The Stokes-flow hydrodynamics of an irregular microparticle is encoded by
its grand resistance matrix --- a $6\!\times\!6$ tensor whose
translational and rotational blocks ($\boldsymbol{A}$ and
$\boldsymbol{C}$) govern settling, diffusion, and orientational transport.
Empirical drag correlations compress these tensors to a single scalar,
discarding drag's orientation dependence and the rotational response
entirely. We present an $\mathrm{SO}(3)$-equivariant neural network that predicts
the full symmetric positive-definite $\boldsymbol{A}$ and $\boldsymbol{C}$
blocks directly from a particle's spherical-harmonic surface
representation, equivariant by construction to floating-point precision.
Trained on $1.1\!\times\!10^{5}$ random shapes from near-spherical to
very rough and tested on a sealed set of $18\,000$, it achieves
$1.4\,\%$ and $2.7\,\%$ mean relative error on the two blocks while
evaluating each shape four to five orders of magnitude faster than the
regularised-Stokeslet solver that generated its labels. A
spectral-convergence study confirms the representation is itself faithful:
truncating a shape at spherical-harmonic degree $\ell=15$ changes its
resistance by a median of only $0.14\,\%$ (translation) and $0.38\,\%$
(rotation), while the training shapes --- generated band-limited at $\ell\le 15$ --- carry no truncation error, so the representation does not limit the surrogate's accuracy. Across
$1.16\!\times\!10^{6}$ orientation-sampled settling, rotation, and
diffusion events, the surrogate reveals lateral drift up to $11^{\circ}$, rotational
misalignment up to $46^{\circ}$, and shape-induced diffusion spreads of
$30\,\%$ (translational) and $2.4\times$ (rotational) --- all identically
zero under any scalar or spheroid reduction. Even the orientation-averaged
scalar friction the correlations target, accurate to $1.7$--$2.8\,\%$
(median; $2.2$--$3.2\,\%$ mean),
carries no tensor orientation, whereas the surrogate reproduces that
scalar to $\sim\!1\,\%$ while supplying the full anisotropic tensors. A fast, equivariant tensor surrogate can
replace both solver and scalar approximation across atmospheric dust
transport, microplastic fate, and colloidal Brownian dynamics.%
} \\

\end{tabular}

\end{@twocolumnfalse} \vspace{0.6cm}]

\renewcommand*\rmdefault{bch}\normalfont\upshape
\rmfamily
\section*{}
\vspace{-1cm}

\footnotetext{\textit{$^{a}$~Department of Mechanical and Aerospace Engineering, University of California, Los Angeles, CA, USA. E-mail: jdeldre@ucla.edu}}
\footnotetext{\textit{$^{b}$~School of Biological and Health Systems Engineering, Arizona State University, Tempe, AZ, USA.}}
\footnotetext{\textit{$^{c}$~Fielding School of Public Health, University of California, Los Angeles, CA, USA. E-mail: candacetsai@ucla.edu}}
\footnotetext{\dag~Electronic Supplementary Information (ESI) available. See DOI: 00.0000/00000000.}

\section{Introduction}\label{sec:intro}

The hydrodynamic transport of micron-scale rigid particles in the viscous-dominated regime governs a remarkable breadth of natural and engineered phenomena. In the atmosphere, the long-range dispersion and deposition of mineral dust aerosols\cite{Kok2011} and of microplastic fragments and fibres\cite{Brahney2020} set the boundary conditions for biogeochemical cycles, radiative forcing, and human exposure budgets. In the human respiratory tract, the settling, diffusion and orientational drift of inhaled pollutants and of pharmaceutical aerosols determine regional deposition patterns and drug-delivery efficacy\cite{Darquenne2012}. In industrial powder handling, fluidised-bed reactors and filtration units, the anisotropic drag and rotational diffusion of irregularly shaped grains control residence times, mixing rates and separation performance\cite{Zhong2016}. In biophysics and colloidal science, the mobility of a rigid body in a quiescent solvent is the central component of Brownian-dynamics simulation, analytical ultracentrifugation and single-molecule experiments\cite{GarciaDeLaTorre1981,Carrasco1999}. Across all of these applications the particle Reynolds number satisfies $\mathrm{Re}_p \ll 1$, the quasi-static Stokes equations apply, and the complete hydrodynamic response of each particle is encoded in a compact, shape-dependent linear object: the $6\times 6$ grand resistance matrix (GRM).

The grand resistance matrix $\boldsymbol{\mathcal{R}}$ relates the rigid-body translational and angular velocities $(\boldsymbol{U},\boldsymbol{\Omega})$ of a particle to the hydrodynamic force $\boldsymbol{F}$ and torque $\boldsymbol{T}$ that the surrounding fluid exerts on the particle through
\begin{equation}
\begin{pmatrix}\boldsymbol{F}\\ \boldsymbol{T}\end{pmatrix}
\;=\; -\,\mu\,
\begin{pmatrix}\boldsymbol{A} & \boldsymbol{B}^{\top} \\[2pt] \boldsymbol{B} & \boldsymbol{C}\end{pmatrix}
\begin{pmatrix}\boldsymbol{U}\\ \boldsymbol{\Omega}\end{pmatrix},
\label{eq:grm}
\end{equation}
where $\mu$ is the fluid viscosity and $\boldsymbol{A}$, $\boldsymbol{B}$, $\boldsymbol{C}$ are $3\times 3$ blocks describing, respectively, the translation--translation, translation--rotation, and rotation--rotation resistances\cite{Brenner1963,HappelBrenner1983,KimKarrila1991}. With the viscosity factored out in eqn~\eqref{eq:grm}, $\boldsymbol{A}$, $\boldsymbol{B}$ and $\boldsymbol{C}$ carry dimensions of length, length squared and length cubed. As a reference point, the Stokes drag on a translating sphere of radius $a$ is $\boldsymbol{F}=-\mu\boldsymbol{A}\boldsymbol{U}$ with $\boldsymbol{A}=6\pi a\,\boldsymbol{I}$, where $\boldsymbol{I}$ is the identity matrix. All numerical resistances in this study are quoted in this convention (equivalently, in units with $\mu=1$) for volume-normalised shapes. The diagonal blocks $\boldsymbol{A}$ and $\boldsymbol{C}$ are symmetric positive-definite and depend only on the particle's shape through an intrinsic polyadic relation first made explicit by Brenner\cite{Brenner1963,Brenner1964b}. The off-diagonal coupling block $\boldsymbol{B}$ is a pseudotensor: it changes sign under spatial inversion, so it vanishes identically for any particle possessing a centre of inversion, and more generally for bodies whose symmetry group is rich enough to eliminate every component---bodies of revolution and orthotropic bodies with three mutually perpendicular mirror planes among them\cite{HappelBrenner1983,KimKarrila1991}. A single mirror plane, by contrast, forces only part of $\boldsymbol{B}$ to zero, so achirality alone does not guarantee a vanishing coupling; chirality is sufficient, but not necessary, for $\boldsymbol{B}\neq\boldsymbol{0}$. For the overwhelming majority of naturally occurring and industrially produced microparticles---fragments of brittle materials released by weathering and abrasion, droplets and crystallites condensed from the vapour phase, atomised metal powders, microplastic debris and fibres, and airborne mineral dust---the formation pathway carries no preferred handedness, $\boldsymbol{B}$ vanishes in the ensemble mean, and its magnitude is small for individual realisations. The Einstein--Smoluchowski relation $\boldsymbol{D} = k_{\mathrm B} T\,(\mu\boldsymbol{\mathcal{R}})^{-1}$ then expresses the translational and rotational diffusion tensors and the orientational relaxation time in terms of the parity-even blocks $\boldsymbol{A}$ and $\boldsymbol{C}$ alone, and the same two blocks fix the sedimentation velocity through the mobility. Predicting these two blocks as a function of shape is the central problem of single-particle microhydrodynamics and a prerequisite for quantitative transport modelling across essentially every environmental, biomedical and industrial application listed above.

Closed-form expressions for the resistance blocks exist only for a handful of idealised shapes: the sphere\cite{Stokes1851}, the spheroid\cite{Perrin1934}, the triaxial ellipsoid\cite{Oberbeck1876,Jeffery1922}, and a small number of other high-symmetry bodies catalogued by Happel and Brenner\cite{HappelBrenner1983} and by Kim and Karrila\cite{KimKarrila1991}. For arbitrary shapes one must resort to numerical solution of the exterior Stokes problem. The boundary integral method (BIM), which reformulates the problem as an integral equation over the particle surface, has become the standard high-fidelity approach since the seminal work of Youngren and Acrivos\cite{YoungrenAcrivos1975}, with the completed-double-layer formulation of Power and Miranda\cite{PowerMiranda1987} providing a second-kind integral equation that is well-posed for particles with closed (watertight) surfaces and remains the basis of many modern implementations\cite{Pozrikidis1992}. The method of regularised Stokeslets\cite{Cortez2001,Cortez2005} replaces point singularities with smooth blobs, eliminating the difficulties of near-singular integration and enabling simulations with complex or deforming geometries. Bead and bead-shell models tailored to biomolecular inputs\cite{GarciaDeLaTorre1981,Carrasco1999} and precise boundary-element implementations\cite{AragonHahn2006} achieve a few percent error on translational and rotational diffusion tensors across benchmark protein datasets, while fast convex-hull surrogates such as HullRad\cite{Fleming2018} trade accuracy for throughput on disordered macromolecular ensembles. Despite decades of algorithmic progress, the cost of these methods scales as $\mathcal{O}(N^2)$--$\mathcal{O}(N^3)$ with the number of surface elements, and a single high-accuracy evaluation takes seconds to minutes of wall time on modern hardware. This cost precludes their direct use in applications that require ensemble statistics over $10^4$--$10^6$ realistic particles: the calibration of atmospheric-transport parameterisations, the tuning of computational-fluid-dynamics--discrete-element-method (CFD--DEM) closures for industrial-scale simulations, the construction of priors for data assimilation, and population-level studies of settling-velocity distributions in aerosols and sediments.

The simulation community has responded by compressing the shape dependence of the resistance matrix to a small set of scalar descriptors and fitting empirical correlations against experimental and direct numerical simulation data. Ganser\cite{Ganser1993} proposed a rational form for the drag coefficient of non-spherical particles based on sphericity and a crosswise projected-area ratio; H\"olzer and Sommerfeld\cite{Holzer2008} introduced a widely adopted correlation employing crosswise and lengthwise sphericities that remains accurate up to the critical Reynolds number and is now a standard closure in CFD--DEM codes. Dioguardi \emph{et al.}\cite{Dioguardi2018} formulated a single-equation drag model for irregular volcanic particles valid across a wide Reynolds range, and Bagheri and Bonadonna\cite{Bagheri2016} calibrated a general drag model against 300 regular and irregular particles freely falling in air together with a compilation of 881 published measurements in liquids, achieving roughly 10\% mean error across the dataset. The reviews of Loth\cite{Loth2008} and, more recently, Michaelides and Feng\cite{MichaelidesFeng2023} summarise the state of the art for drag on solid particles of regular and irregular shape. These correlations have been enormously useful, but they share a fundamental limitation: each compresses the anisotropic tensor $\boldsymbol{A}$ to a single scalar, discards the rotation tensor $\boldsymbol{C}$ entirely, and retains no information about how the principal axes of the resistance tensors are oriented relative to the particle body frame. Their predictions are necessarily orientation-averaged and cannot reproduce, for example, the order-of-magnitude anisotropy in settling velocity between a long microplastic fibre falling broadside-on versus end-on, or the rotational Brownian relaxation of an asymmetric dust grain whose three principal rotational friction coefficients can differ several-fold.

Machine learning has recently been enlisted to push these scalar correlations further, part of a broader turn to data-driven modelling in fluid mechanics\cite{Brunton2020}. Hwang, Pan and Fan\cite{Hwang2021} combined a variational autoencoder trained on spherical-harmonic shape coefficients with a multilayer perceptron to predict drag, lift and pitching torque on non-spherical and irregular particles in low-Reynolds-number incompressible flow. Subsequent work has extended neural drag prediction to broader shape families\cite{Xiang2024}, to the many-body drag of dense particle assemblies for use as closures in CFD--DEM\cite{HeTafti2019} and to polydisperse assemblies of irregular grains\cite{Hwang2024}, and to physics-informed training that embeds the classical correlations directly in the network\cite{PresaReyes2024}. In parallel, the HIGNN framework of Ma \emph{et al.}\cite{Ma2022} and its descendants use graph neural networks to learn many-body corrections on top of the analytically known single-sphere mobility, enabling fast simulation of Stokes suspensions at scales inaccessible to direct BIM. Across this entire line of research, however, the output of the network is either (i)~a scalar drag coefficient or a small collection of scalar force and torque components evaluated at a fixed orientation, or (ii)~a many-body correction that presupposes single-particle mobility is analytically known. None of the existing methods predicts the full anisotropic $3\times 3$ blocks $\boldsymbol{A}$ and $\boldsymbol{C}$ for arbitrary single-particle shapes, which is the minimum information required to describe orientation-resolved drag, rotational diffusion, and coupled translation--orientation dynamics of a non-spherical grain.

This scalar ceiling is structural. A neural network that predicts a shape-dependent tensor must respect the symmetry of the underlying physics: a rigid rotation $\boldsymbol{R}\in\mathrm{SO}(3)$ (the group of rotations in three dimensions) applied to the particle should induce the corresponding conjugation of the output, $\boldsymbol{A}\mapsto\boldsymbol{R}\boldsymbol{A}\boldsymbol{R}^{\top}$, without learning this transformation law from data. Architectures that ingest rotation-invariant scalar descriptors cannot produce outputs that transform as a rank-two tensor, and architectures operating on raw point clouds or meshes must fall back on data augmentation, which scales poorly and cannot guarantee exact equivariance. The last decade has, however, seen the emergence of group-equivariant neural networks---networks whose outputs transform in concert with rotations of the input---that build these symmetry constraints into their weight structure by construction\cite{Cohen2016,Weiler2018}. Tensor field networks, introduced by Thomas \emph{et al.}\cite{Thomas2018}, operate on three-dimensional point clouds through spherical-harmonic filters and produce outputs that are exactly equivariant under $\mathrm{SO}(3)$ rotations and $\mathbb{R}^3$ translations, naturally handling scalars, vectors, and higher-order tensors as first-class objects in the same architecture. The \texttt{e3nn} library of Geiger and Smidt\cite{Geiger2022} has made these architectures widely accessible and triggered a wave of applications in atomistic modelling, where the NequIP potential of Batzner \emph{et al.}\cite{Batzner2022} established that $E(3)$-equivariant graph neural networks reach a given accuracy with up to three orders of magnitude less training data than non-equivariant counterparts while producing force fields that transform correctly under rotation. Equivariant machinery has transformed molecular property prediction, interatomic potentials, and catalyst design, but---to the best of our knowledge---it has not previously been applied to the shape-to-resistance problem that is the foundational object of microhydrodynamics.

We hypothesise that this is precisely the problem where equivariant neural networks should excel. The output of interest is a pair of symmetric positive-definite $3\times 3$ matrices whose components transform as a known representation of $\mathrm{SO}(3)$: each of $\boldsymbol{A}$ and $\boldsymbol{C}$ decomposes irreducibly into a scalar trace ($\ell=0$) and a traceless symmetric rank-two part ($\ell=2$). When the particle shape is represented by the coefficients of its spherical-harmonic expansion---a representation that has become a standard tool for the morphological analysis of concrete aggregates, sand grains, volcanic ash and other natural particles since the X-ray tomography work of Garboczi\cite{Garboczi2002}---the input is indexed by the same $\mathrm{SO}(3)$ irreps, and the entire map from input to output sits cleanly inside the equivariant category. Because both target tensors are parity-even, the network needs only parity-even output heads---no parity-odd output is required---which keeps the surrogate simple, fast, and data-efficient for the statistically achiral regime in which nearly all natural and industrial microparticles reside. The chirality-sensitive coupling block $\boldsymbol{B}$, by contrast, is a pseudotensor whose prediction would demand a parity-odd output head and which we accordingly defer. The diverse training distribution required to generalise across real morphologies is generated from the Cartesian spherical-harmonic (Legendre-block) parameterisation, calibrated against natural-particle statistics and extended into the rougher, more elongated regimes of weathered aggregates and microplastics. Tensor labels are produced at scale by a standard six-shot regularised-Stokeslet-surface Stokes solver with analytic per-triangle integration, whose accuracy we benchmark against analytical sphere and spheroid solutions.

In this work we present what is, to the best of our knowledge, the first equivariant neural surrogate for the full anisotropic translational and rotational resistance tensors of arbitrarily shaped microparticles in Stokes flow. Our specific contributions are as follows.
\begin{enumerate}
\item[(i)] \textit{Tensor output.}\; We formulate the shape-to-resistance problem as regression of the symmetric positive-definite blocks $\boldsymbol{A}$ and $\boldsymbol{C}$ of the grand resistance matrix, decomposing each into its $\ell=0$ and $\ell=2$ irreducible components so that the output transforms exactly under rigid rotations of the input.
\item[(ii)] \textit{Equivariant architecture.}\; We deploy SHEAR (Spherical-Harmonic Equivariant Architecture for Resistance), an $\mathrm{SO}(3)$-equivariant network acting directly on the spherical-harmonic shape coefficients, implemented in \texttt{e3nn}\cite{Geiger2022,Batzner2022} from Clebsch--Gordan tensor products and gated nonlinearities with a compact mixed-parity hidden basis and log-Euclidean symmetric-positive-definite output heads; it carries no hand-computed invariant side-channel. To our knowledge this is the first architecture to produce orientation-resolved hydrodynamic resistance tensors at quantitative accuracy across a heterogeneous population of shapes.
\item[(iii)] \textit{Large, realistic training set.}\; We generate $\mathcal{O}(10^5)$ training shapes from the calibrated Cartesian spherical-harmonic parameterisation, spanning near-spherical, ellipsoidal and rough-aggregate regions of morphology space, with shape statistics consistent with published measurements of mineral dust\cite{Huang2020}, volcanic ash\cite{Bagheri2016}, and microplastic fragments\cite{Brahney2020}.
\item[(iv)] \textit{Reference-quality labels.}\; All training labels are produced by a standard six-shot Stokes solver based on the method of regularised Stokeslet surfaces, with analytic integration of the regularised kernel over each surface triangle. The solver achieves sub-percent error on the analytical sphere and spheroid resistance tensors while remaining tractable on GPU hardware at dataset scale.
\item[(v)] \textit{Speed, accuracy, and scope.}\; The trained surrogate evaluates the full $(\boldsymbol{A},\boldsymbol{C})$ pair in tens of microseconds per shape on commodity GPU hardware ($57\,\mu$s at batch $10^4$), four to five orders of magnitude faster than the regularised-Stokeslet-surface solver. We benchmark against the classical correlations of Ganser\cite{Ganser1993}, H\"olzer and Sommerfeld\cite{Holzer2008} and Bagheri and Bonadonna\cite{Bagheri2016}, and against a non-equivariant control---a multilayer perceptron that regresses the resistance eigenvalues from rotation-invariant shape descriptors---which isolates the contribution of equivariance, and we delimit the scope of the present parity-even output formulation, which targets the statistically achiral regime of natural microparticles.
\end{enumerate}

The remainder of this study is organised as follows.
Section~\ref{sec:methods} describes the spherical-harmonic shape
representation and sampling, the mesh construction and canonicalisation,
and the regularised-Stokeslet solver that generates the resistance labels,
along with the methods specific to each result. Section~\ref{sec:results}
reports the three results: how faithfully a particle's shape must be
resolved before its resistance is fixed (Result~1), how accurately and
how fast the equivariant surrogate predicts that resistance
(Result~2), and what the resulting tensors imply for the settling,
rotation and diffusion that no scalar reduction can reproduce
(Result~3).
Section~\ref{sec:discussion} examines scope and limitations, and
Section~\ref{sec:conclusions} concludes.

\section{Methodology}\label{sec:methods}

The pipeline has three sequential steps. We first specify the parametric shape
representation and the stochastic process that draws training samples
from it (Section~\ref{sec:shapes}). We then construct a surface
triangulation of each shape and place it in a common reference frame
(Section~\ref{sec:canon}). Finally, we compute the grand resistance
matrix for every shape with a Stokes solver based on the method of
regularised Stokeslet surfaces (Section~\ref{sec:solver}).
Section~\ref{sec:spectral_methods} then quantifies how faithfully the
truncated spherical-harmonic representation captures the resistance,
and Section~\ref{sec:irrep} sets out the irreducible tensor structure
of the grand resistance matrix around which the equivariant surrogate
of Section~\ref{sec:model} is built.

\begin{figure*}[t]
\centering
\includegraphics[width=\textwidth]{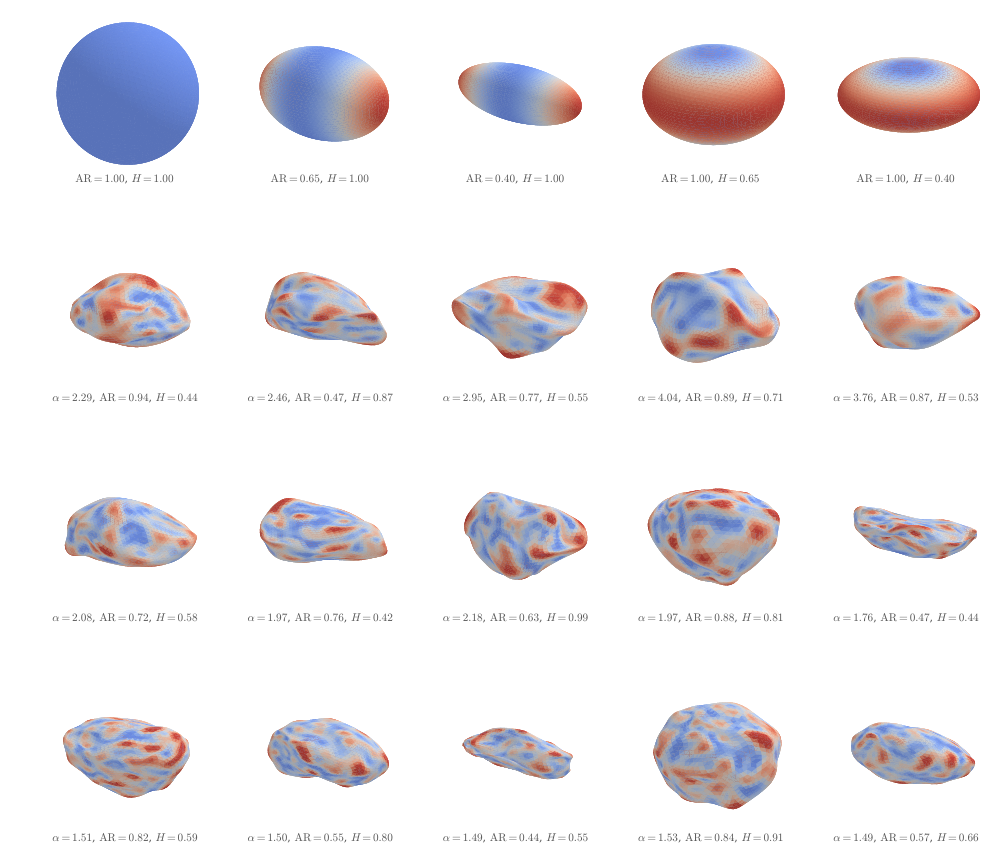}
\caption{Twenty representative particles drawn from the sealed test
  set, spanning the morphology window of
  Section~\ref{sec:shapes}. The top row shows near-ellipsoidal
  references; the lower rows are ordered by increasing surface
  roughness (i.e.\ decreasing fitted spectral exponent
  $\alpha_{\rm fit}$), from smooth pebble-like grains down to rough
  multi-scale aggregates. Columns are
  ordered by Zingg morphological class\cite{Zingg1935}:
  near-spherical, prolate (elongated), oblate (flat), triaxial,
  and an extreme elongated-flat combination. Each panel is annotated
  with the shape's fitted $\alpha_{\rm fit}$ together with the
  first-degree-ellipsoid aspect ratio $\mathrm{AR}=p_2/p_1$ and
  flatness $H=p_3/p_2$. Renders use a coolwarm gradient along the
  longest principal axis modulated by Lambertian shading; meshes
  are at icosphere subdivision $r=3$ ($V=642$, $F=1280$) for
  legibility at print size, although every shape was meshed at
  $r=4$ for the resistance-tensor solve of
  Section~\ref{sec:solver}.}
\label{fig:gallery}
\end{figure*}

\subsection{Shape parameterisation and sampling}\label{sec:shapes}

No single morphology class serves all of the applications targeted here.
The training ensemble is therefore drawn from three overlapping families
within a single parameterisation --- smooth ellipsoids, natural sand
grains, and strongly roughened aggregates --- spanning the idealised
shapes for which analytic resistances are known, the measured statistics
of natural sediment, and the rougher, more elongated regimes of weathered
aggregates and microplastics; the three are pooled and split into the
training and sealed test sets used throughout.

We represent each particle surface as a vector-valued map
$\boldsymbol{s}:S^2\to\mathbb{R}^3$ expanded in real orthonormal
spherical harmonics (SH),
\begin{equation}
\boldsymbol{s}(\theta,\varphi)
= \sum_{\ell=0}^{L_{\max}}\sum_{m=-\ell}^{\ell}
    \boldsymbol{\mathsf{C}}_\ell^m\,Y_\ell^m(\theta,\varphi),\qquad
\boldsymbol{\mathsf{C}}_\ell^m\in\mathbb{R}^3,
\label{eq:lbs}
\end{equation}
where \(\ell\) is the harmonic degree, \(m\in\{-\ell,\ldots,\ell\}\) its
order, and \(L_{\max}\) the truncation degree. The real spherical
harmonics satisfy the orthonormality relation
$\int_{S^2}Y_\ell^m Y_{\ell'}^{m'}\mathrm{d}\Omega
=\delta_{\ell\ell'}\delta_{mm'}$, and each \((\ell,m)\) pair is flattened
to a single array index $k(\ell,m)=\ell^2+\ell+m$, so that the
$(L_{\max}+1)^2$ coefficients up to degree $L_{\max}$ occupy the
contiguous range $k=0,\ldots,(L_{\max}+1)^2-1$. The three Cartesian
channels $(x,y,z)$ give a
real coefficient matrix
$\boldsymbol{\mathsf{C}}\in\mathbb{R}^{3\times(L_{\max}+1)^2}$. This
Cartesian spherical-harmonic (CSH) representation was introduced by
Wei, Wang and Zhao\cite{Wei2018}; unlike the radial-scalar expansion
$r(\theta,\varphi)=\sum c_\ell^m Y_\ell^m$, it accommodates non-star
surfaces, whose radius is not a single-valued function of direction
from any interior point (concavities, overhangs), because each
coordinate is expanded independently, and it cleanly separates the
low-degree ellipsoidal skeleton from higher-degree roughness. We
truncate the training expansion at $L_{\max}=15$ unless stated otherwise
($256$ coefficients per channel, $768$ real numbers per shape); the
spectral-convergence study of Section~\ref{sec:spectral_methods} uses
$L_{\max}=40$ as its reference and reports truncations up to
degree $L=30$.

\paragraph{First-degree ellipsoid.}
Setting all coefficients with $\ell\neq 1$ to zero, eqn~\eqref{eq:lbs}
reduces to the axis-aligned ellipsoid with semi-axes
$a\geq b\geq c>0$ through
\begin{equation}
\boldsymbol{\mathsf{C}}_1
= \sqrt{\frac{4\pi}{3}}
\begin{pmatrix}
0 & 0 & -a \\
-b & 0 & 0 \\
0 & +c & 0
\end{pmatrix},
\label{eq:C1_block}
\end{equation}
where the columns correspond to $m\in\{-1,0,+1\}$ in the sign
convention of Wei et~al.\cite{Wei2018}. We call this the
\emph{first-degree ellipsoid} (FDE) and parameterise it by the aspect
ratio $\mathrm{AR}=b/a\in(0,1]$ and flatness $H=c/b\in(0,1]$, fixing
$a=1$ so the semi-major axis is unity before volume normalisation
(Section~\ref{sec:canon}).

\paragraph{Two-segment roughness spectrum.}
For each $\ell\geq 2$ the rotation-invariant descriptor
$d_\ell=\|\boldsymbol{\mathsf{C}}_\ell\|_F$ measures the surface energy carried
at angular wavenumber~$\ell$. Wei, Wang and Zhao\cite{Wei2018},
building on the rotation-invariant spherical-harmonic analysis of
Zhao, Wei and Wang\cite{Zhao2017}, observed empirically (their Eq.~5)
that natural sand and ash grains
do not follow a single power law. Instead the spectrum splits at
$\ell=9$ into a low-degree \emph{roundness} band
($\ell\in[2,8]$) and a steeper high-degree \emph{roughness} band
($\ell\in[9,15]$), with negligible energy at $\ell>15$ in their
fitted dataset. We adopt this two-segment form. Let
$D_{2{-}8} =\sum_{\ell=2}^{8}d_\ell/d_1$ and
$D_{9{-}15}=\sum_{\ell=9}^{15}d_\ell/d_1$ denote the band-integrated
energies normalised by $d_1=\|\boldsymbol{\mathsf{C}}_1\|_F$. Within each band
the per-degree weights decay as a power law,
\begin{equation}
d_\ell \;=\;
\begin{cases}
\displaystyle d_1\,D_{2{-}8}\,\frac{(\ell/2)^{-\alpha}}
                                {\sum_{\ell'=2}^{8}(\ell'/2)^{-\alpha}},
& 2\leq\ell\leq 8,\\[10pt]
\displaystyle d_1\,D_{9{-}15}\,\frac{(\ell/9)^{-\beta}}
                                {\sum_{\ell'=9}^{15}(\ell'/9)^{-\beta}},
& 9\leq\ell\leq 15,
\end{cases}
\label{eq:wei_two_segment}
\end{equation}
which preserves the band integrals exactly while distributing energy
within each band by the slope exponents
$(\alpha,\beta)\in\mathbb{R}_{>0}^{2}$. Fitting eqn~\eqref{eq:wei_two_segment}
to the 80-grain Leighton Buzzard sand dataset of
Wei~et~al.\cite{Wei2018,Zhao2017} yields anchor values $\alpha=1.387$,
$\beta=1.426$, with the original paper's confidence intervals
$\alpha\in[1.32,1.45]$ and $\beta\in[1.19,1.67]$. Natural sand sits at
$D_{2{-}8}\approx 0.10$--$0.30$ and $D_{9{-}15}\lesssim 0.06$. The
conventional single-slope roughness spectrum
$d_\ell\propto(2/\ell)^{\alpha}$---a single power law across all
degrees, as used in earlier spherical-harmonic shape models before the
two-band split of Wei et~al., and also the form we adopt for the
sphere-baseline ensemble of the spectral-convergence study
(Section~\ref{sec:spectral_methods})---is recovered from
eqn~\eqref{eq:wei_two_segment} by setting $\alpha=\beta$ and choosing
the two band energies so that the per-degree weights $d_\ell$ join
smoothly across the $\ell=9$ split into one unbroken power law (which
fixes $D_{9{-}15}/D_{2{-}8}$ to the ratio of the two bands' summed
weights). Each
degree-$\ell$ block is then realised as a $3\times(2\ell+1)$ Gaussian
random matrix rescaled to the target Frobenius norm,
\begin{equation}
\boldsymbol{\mathsf{C}}_\ell
= d_\ell\,\frac{\boldsymbol{G}_\ell}{\|\boldsymbol{G}_\ell\|_F},\qquad
(\boldsymbol{G}_\ell)_{ij}\sim\mathcal{N}(0,1),
\label{eq:random_Cl}
\end{equation}
which places $\boldsymbol{\mathsf{C}}_\ell$ at a uniformly random point on the
sphere of radius $d_\ell$ in the space $\mathbb{R}^{3(2\ell+1)}$ of
degree-$\ell$ coefficient blocks (three Cartesian channels $\times$
$2\ell+1$ orders): the direction is isotropic and the magnitude is
fixed to $d_\ell$. The $\ell=0$ coefficient is set to zero, and the
particle's centroid is fixed at the canonicalisation stage of
Section~\ref{sec:canon}.

\paragraph{Beyond the natural-sand window.}
The natural sand and ash ranges of Wei et~al.\cite{Wei2018} cover
only a fraction of the morphology space relevant to the applications
that motivate this work. Microplastic fragments produced by mechanical
abrasion are rougher and more elongated than river sand; volcanic
ash carries a finer, more bubble-like surface texture than smoothed
sediment\cite{Bagheri2016}; and mineral-dust aggregates exhibit a
genuine multi-scale roughness band between the resolution of $\mu$CT
imaging and that of finer surface-texture measurements\cite{Yang2016,WeiFractal2018}
that the natural-window $D_{9{-}15}$
substantially under-represents. We therefore train and test on a
\emph{deliberately extended} window,
\begin{equation}
\begin{aligned}
\mathrm{AR}    &\in [0.25,\,1.0], &
H              &\in [0.20,\,1.0], \\
D_{2{-}8}      &\in [0.0,\,0.70],  &
D_{9{-}15}     &\in [0.0,\,0.15],
\end{aligned}
\label{eq:wei_extended_ranges}
\end{equation}
with the slope exponents fixed at the natural-sand anchors
$\alpha=1.387$ and $\beta=1.426$. The window retains the natural ranges
of Wei et~al.\ as a sub-box and extends the aspect-ratio, flatness and
band-energy parameters by factors of $\sim\!2$--$3$ into regimes
populated by industrially produced microplastics, weathered aggregates
and surface-textured ash. It is covered by three overlapping families
--- smooth ellipsoids ($D_{2{-}8}=D_{9{-}15}=0$), a natural-window sand
family ($D_{2{-}8}\leq 0.5$, $D_{9{-}15}\leq 0.06$), and an extended
rough family ($D_{2{-}8}\in[0.3,\,0.7]$,
$D_{9{-}15}\in[0.04,\,0.15]$) --- giving a training distribution that
spans near-spherical grains through very rough aggregates.

\begin{figure*}[t]
  \centering
  \begin{subfigure}[t]{0.32\textwidth}
    \centering
    \includegraphics[width=\linewidth]{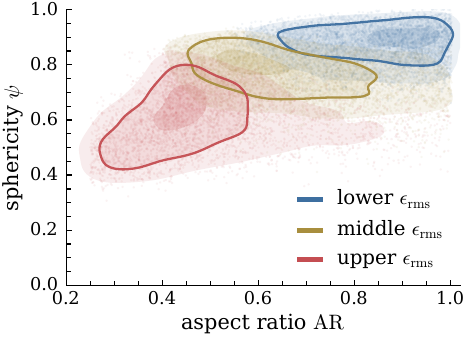}
    \caption{Morphological coverage.}
    \label{fig:diversity_geometry}
  \end{subfigure}\hfill
  \begin{subfigure}[t]{0.32\textwidth}
    \centering
    \includegraphics[width=\linewidth]{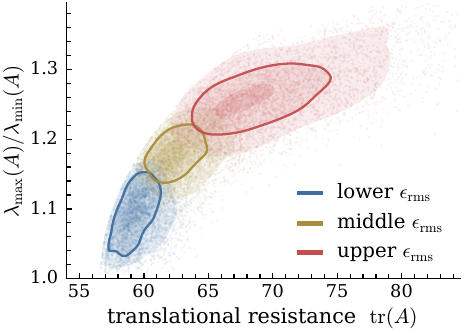}
    \caption{Translational resistance.}
    \label{fig:diversity_A}
  \end{subfigure}\hfill
  \begin{subfigure}[t]{0.32\textwidth}
    \centering
    \includegraphics[width=\linewidth]{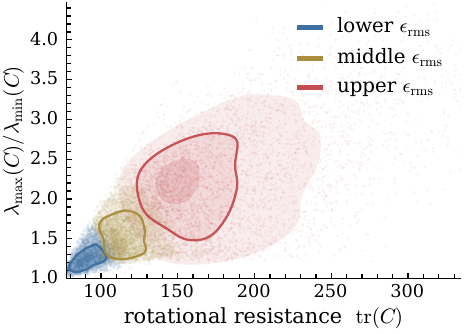}
    \caption{Rotational resistance.}
    \label{fig:diversity_C}
  \end{subfigure}
  \caption{\textbf{Diversity of the training ensemble} (sealed test set),
    coloured by RMS-roughness tercile \(\varepsilon_{\rm rms}\)
    (lower/middle/upper). In each panel the faint dots are individual
    particles, the shaded blobs are nested Gaussian kernel-density levels
    (\(20/50/85\%\) of each tercile's peak), and the solid curve is the
    \(50\%\) contour, marking where each tercile concentrates; axes are
    clipped at the \(99^{\rm th}\) percentile.
    \emph{(a)}~Shape coverage --- aspect ratio \(\mathrm{AR}\) versus
    sphericity \(\psi\) --- spanning near-spherical smooth grains to
    elongated, low-sphericity aggregates.
    \emph{(b)}~Translational resistance, \(\mathrm{tr}(\boldsymbol{A})\)
    versus anisotropy
    \(\lambda_{\max}(\boldsymbol{A})/\lambda_{\min}(\boldsymbol{A})\).
    \emph{(c)}~Rotational resistance, \(\mathrm{tr}(\boldsymbol{C})\)
    versus anisotropy, whose ratio spans nearly fourfold (rotational
    friction scales with the cube of the body size). Rougher grains
    carry higher mean drag and stronger anisotropy in both blocks.}
  \label{fig:diversity}
\end{figure*}

Figure~\ref{fig:gallery} shows a
representative selection from the resulting ensemble, and
Fig.~\ref{fig:diversity} quantifies its diversity in both morphology and
resistance-tensor space.

\subsection{Mesh construction and canonicalisation}\label{sec:canon}

For every sampled $\boldsymbol{\mathsf{C}}$ we build a surface triangulation by
evaluating eqn~\eqref{eq:lbs} on the vertices of a unit icosphere at
subdivision level $r$ (giving $V=10\cdot 4^r+2$ vertices and
$F=20\cdot 4^r$ faces). The parametric angles
$(\theta_i,\varphi_i)$ of each unit-sphere vertex are fixed, and the
corresponding three-dimensional position is
$\boldsymbol{v}_i=\sum_k Y_k(\theta_i,\varphi_i)\,\boldsymbol{\mathsf{C}}_{:,k}$
through the spherical-harmonic design matrix. Throughout this work we
use $r=4$, yielding $V=2562$ and $F=5120$.

Because a deformed CSH surface generally has a spatially varying Jacobian,
the face areas of the mesh evaluated on a uniform icosphere can differ
by more than an order of magnitude across the surface, which
deteriorates the conditioning of the discretised Stokes system. We
therefore improve the mesh with a centroidal Voronoi
tessellation\cite{Lloyd1982,Du1999} performed entirely in the
sphere-parameter domain and weighted by the three-dimensional face
areas: at each iteration every sphere vertex is moved toward the
area-weighted centroid of its one-ring neighbourhood, re-projected
onto the unit sphere, and the three-dimensional positions are
recomputed. Convergence is declared when the coefficient of variation
of the face areas satisfies $\mathrm{CV}(A)<0.10$. Because the
procedure never assumes that the surface is star-shaped from the
origin, it handles folded and self-intersecting CSH surfaces without
modification.

Each converged mesh is then placed in a common reference frame by
subtracting its area-weighted surface centroid and scaling
isotropically to the unit-sphere volume $V_0=4\pi/3$,
\begin{equation}
\boldsymbol{v}_i
\;\leftarrow\;
\left(\frac{V_0}{V_{\mathrm{mesh}}}\right)^{\!1/3}\!
\bigl(\boldsymbol{v}_i-\boldsymbol{x}_{c}\bigr),
\label{eq:canonicalisation}
\end{equation}
where
$\boldsymbol{x}_{c}=\sum_k A_k\bar{\boldsymbol{v}}_k/\sum_k A_k$ is the
area-weighted face centroid and $V_{\mathrm{mesh}}$ is the signed
volume of the triangulation\cite{Zhang2001}. Any fixed, shape-intrinsic
reference point serves equally well here; we adopt the area-weighted
\emph{surface} centroid because it is inexpensive to compute on the
triangulation, and the solver then evaluates torques about the same
point. Using one reference point throughout ensures that the
spherical-harmonic coefficients and the resistance labels are defined
in the same frame.

\subsection{Regularised Stokeslet surface solver}\label{sec:solver}

We solve the exterior Stokes problem for every particle by the method
of regularised Stokeslet surfaces\cite{Ferranti2024}, which replaces
the singular Oseen--Burgers tensor with its
$\varepsilon$-regularised blob\cite{Cortez2001,Cortez2005} and
integrates the resulting kernel \emph{analytically} over each surface
triangle using the piecewise-linear force density as the only
approximation. Because the per-triangle integrals are closed-form
rather than quadrature-based, the method converges as
$\mathcal{O}(h^2)$ in the mesh spacing for a fixed regularisation
$\varepsilon$ and does not accumulate near-singular error as vertices
approach one another. The closed-form per-triangle integrals and their
recursions are derived in full by Ferranti and Cortez\cite{Ferranti2024}
and are not reproduced here.

The solver reduces the exterior Stokes problem to a dense linear system
on the mesh. We expand the unknown Stokeslet strength density
$\boldsymbol{g}$---the force per unit area that the particle surface
exerts \emph{on the fluid}, equal and opposite to the traction the
fluid exerts on the particle---in the piecewise-linear nodal
finite-element basis
$\{N_j\}_{j=1}^{V}$ over the $V$ mesh vertices---each basis function
$N_j$ equals one at vertex $j$, zero at every other vertex, and varies
linearly across the triangles in between---so that
$\boldsymbol{g}(\boldsymbol{y})=\sum_j N_j(\boldsymbol{y})\,\boldsymbol{g}_j$,
with $\boldsymbol{g}_j$ the strength density at vertex $j$. The velocity
induced at a field vertex, located at $\boldsymbol{v}_i$, is then the
regularised single-layer integral of this density against the
regularised Stokeslet $\boldsymbol{S}^{\varepsilon}$,
\begin{equation}
\boldsymbol{u}_i=\sum_{j=1}^{V}\boldsymbol{M}^{\varepsilon}_{ij}\,\boldsymbol{g}_j,
\qquad
\boldsymbol{M}^{\varepsilon}_{ij}
=\frac{1}{8\pi\mu}\sum_{T}\int_{T}\boldsymbol{S}^{\varepsilon}(\boldsymbol{v}_i,\boldsymbol{y})\,N_j(\boldsymbol{y})\,\mathrm{d}S(\boldsymbol{y}),
\label{eq:mobility_assembly}
\end{equation}
where $\boldsymbol{u}_i$ is the velocity at vertex $i$,
$\boldsymbol{S}^{\varepsilon}(\boldsymbol{v}_i,\boldsymbol{y})$ is the
regularised Stokeslet kernel between the field point $\boldsymbol{v}_i$
and a source point $\boldsymbol{y}$ on the surface (ESI, Section~S2),
the sum runs over the
mesh triangles $T$, and each per-triangle integral is evaluated in
closed form\cite{Ferranti2024}. Collecting these contributions assembles
the dense mobility matrix
$\boldsymbol{M}_{\varepsilon}\in\mathbb{R}^{3V\times 3V}$, which maps the
$3V$ nodal strength densities to the $3V$ nodal velocities. Assembly
dominates the
per-shape cost, scaling as $\mathcal{O}(F\!\cdot\!V)$ because every
triangle contributes to every field vertex (GPU implementation details
in the ESI, Section~S3).

The grand resistance matrix follows from the standard six-shot
protocol\cite{KimKarrila1991}: a unit translation along each coordinate
axis with no rotation,
$(\boldsymbol{U}_k,\boldsymbol{\Omega}_k)=(\boldsymbol{e}_k,\boldsymbol{0})$
for $k=1,2,3$, and a unit rotation about each axis with no translation,
$(\boldsymbol{U}_k,\boldsymbol{\Omega}_k)=(\boldsymbol{0},\boldsymbol{e}_{k-3})$
for $k=4,5,6$. For each of these six canonical rigid-body motions
we prescribe the no-slip boundary velocity
$\boldsymbol{u}^{(k)}_j=\boldsymbol{U}_k+\boldsymbol{\Omega}_k\times(\boldsymbol{v}_j-\boldsymbol{x}_c)$
and solve for the strength density that produces it,
\begin{equation}
\boldsymbol{M}_{\varepsilon}\,\boldsymbol{g}^{(k)}=\boldsymbol{u}^{(k)},
\qquad k=1,\ldots,6.
\label{eq:sixshot_solve}
\end{equation}
Because all six shots share the \emph{same} operator
$\boldsymbol{M}_{\varepsilon}$, we factorise it once,
$\boldsymbol{M}_{\varepsilon}=\boldsymbol{L}\boldsymbol{U}$, and reuse
that factorisation across the six right-hand sides---each shot then
costs only a pair of triangular solves---so the per-shape cost is
dominated by a single dense factorisation of the $7686\times 7686$
system rather than by six independent solves. Because the fluid
traction on the particle is the negative of the strength density,
integrating each solved density exactly over the triangulation gives
the hydrodynamic force and torque on the particle,
\begin{equation}
\boldsymbol{F}^{(k)}=-\sum_{T}\int_{T}\boldsymbol{g}^{(k)}\,\mathrm{d}S,
\qquad
\boldsymbol{T}^{(k)}=-\sum_{T}\int_{T}(\boldsymbol{y}-\boldsymbol{x}_c)\times\boldsymbol{g}^{(k)}\,\mathrm{d}S,
\label{eq:force_torque}
\end{equation}
and stacking the six force--torque pairs, each obeying
$(\boldsymbol{F}^{(k)},\boldsymbol{T}^{(k)})=-\mu\,\boldsymbol{\mathcal{R}}\,(\boldsymbol{U}_k,\boldsymbol{\Omega}_k)$,
populates the $6\times 6$ matrix $\boldsymbol{\mathcal{R}}$ of
eqn~\eqref{eq:grm}. We assemble and factorise
$\boldsymbol{M}_{\varepsilon}$ on the GPU in double precision. The
full six-shot solve completes in approximately~$1.6$\,s per shape on
an NVIDIA RTX\,5080.

\begin{figure*}[t]
  \centering
  \includegraphics[width=0.94\textwidth]{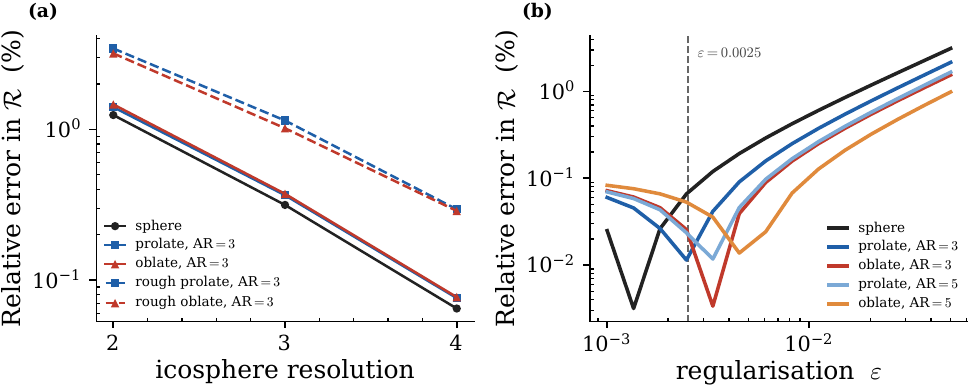}
  \caption{\textbf{Solver-setup convergence at the adopted settings.}
    \emph{(a)} Mesh self-convergence of the full grand-resistance
    matrix, \(\|R_r-R_5\|_F/\|R_5\|_F\), versus icosphere resolution
    at fixed \(\varepsilon=0.005\), for the sphere and smooth
    aspect-ratio-3 prolate/oblate spheroids (solid) and for the same
    spheroids carrying \(10\%\) RMS multiplicative random roughness
    of degree \(\ell\leq 12\) (dashed): resolution~4 agrees with
    resolution~5 to \(<0.08\%\) (smooth) and \(<0.3\%\) (rough).
    \emph{(b)} Relative error of the full matrix against the exact
    analytic resistance versus the regularisation amplitude
    \(\varepsilon\) at resolution~4; the error is minimised near
    \(\varepsilon\approx0.002\)--\(0.004\) (dashed line at \(0.002\)) and the residual
    \(\varepsilon\)-bias is removed by the three-point Richardson
    extrapolation wherever converged reference resistances are
    required (Section~\ref{sec:spectral_methods}); the dataset labels
    use a single solve at \(\varepsilon=0.005\).}
  \label{fig:setup_convergence}
\end{figure*}

Although a fixed $\varepsilon$ at the chosen mesh resolution gives
sub-percent accuracy on analytical benchmarks, the regularisation leaves a residual bias:
the familiar $\mathcal{O}(\varepsilon^{2})$ accuracy of the regularised
Stokeslet holds in the far field\cite{Cortez2001,Cortez2005}, but for
the single-layer kernel evaluated \emph{on} the surface the leading
regularisation error is $\mathcal{O}(\varepsilon)$\cite{GallagherSmith2021}. All dataset labels are
computed with a single six-shot solve per shape at
$\varepsilon=0.005$. From the $\varepsilon$-sweep of
Fig.~\ref{fig:setup_convergence}b this fixed-$\varepsilon$ bias is
$\lesssim 0.1\%$ on the translational block and $\lesssim 0.3\%$ on
the rotational block, is uniform across the dataset, and lies an
order of magnitude below the surrogate's own prediction error
(Result~2). Wherever a \emph{converged} reference resistance is
required---the $L=40$ truth surfaces and the single-mode kernel
measurements of the spectral-convergence study
(Section~\ref{sec:spectral_methods})---we additionally remove this
last source of bias by the sequential Richardson extrapolation in
$\varepsilon$ of Gallagher and Smith\cite{GallagherSmith2021}: we repeat the six-shot solve
at the three coarse values $\varepsilon\in\{0.005,\,0.0075,\,0.01\}$,
fit each entry of the resulting
$\boldsymbol{\mathcal{R}}(\varepsilon)$ exactly by the three-point
ansatz
$\boldsymbol{\mathcal{R}}(\varepsilon)=\boldsymbol{\mathcal{R}}_{0}
+\boldsymbol{\mathcal{R}}_{1}\varepsilon
+\boldsymbol{\mathcal{R}}_{2}\varepsilon^{2}$ --- the linear term
capturing the leading on-surface regularisation bias --- and report
$\boldsymbol{\mathcal{R}}_{0}$ as the converged reference. The three
solves reuse the same mesh and the same vertex layout, so the
marginal cost is only two additional dense factorisations per shape.

Throughout the remainder of this study we fix $\varepsilon=0.005$ at
mesh resolution $r=4$, giving a regularisation-to-spacing ratio
$\varepsilon/h\approx 0.07$. For the analytical benchmark of a
quiescent sphere this configuration recovers
$\tfrac13\operatorname{tr}(\boldsymbol{A})=6\pi a\,(1\pm 10^{-3})$ and
$\tfrac13\operatorname{tr}(\boldsymbol{C})=8\pi a^3\,(1\pm 5\times 10^{-3})$
in the $\mu$-factored convention of eqn~\eqref{eq:grm}.
On the analytical prolate and oblate spheroid
solutions\cite{KimKarrila1991} over aspect ratios from 1:1 to 10:1,
every translational and every rotational coefficient is recovered to
within $0.10\%$ and $0.52\%$ respectively (ESI, Section~S4). An
independent quality check on the output itself: in the achiral FDE
limit ($d_\ell/d_1=0$ for all $\ell\geq 2$) both the symmetry
(reciprocity) violation
$\|\boldsymbol{\mathcal{R}}-\boldsymbol{\mathcal{R}}^{\!\top}\|/\|\boldsymbol{\mathcal{R}}\|$
and the coupling-block magnitude sit at float64 round-off (median
$\sim\!10^{-15}$), as required for an exact, achiral solver; across the
full random dataset the reciprocity violation has median
$\sim\!10^{-5}$, orders of magnitude below every truncation and
surrogate error considered in this study.

Figure~\ref{fig:setup_convergence} confirms that both settings are
converged. Refining the icosphere from resolution~4 (\(2562\) vertices) to~5
(\(10242\) vertices) at fixed \(\varepsilon\) changes the full grand-resistance
matrix by less than \(0.08\%\) for smooth sphere and aspect-ratio-3 spheroid
bases, and by less than \(0.3\%\) for the same bases carrying a \(10\%\) RMS
surface roughness (Fig.~\ref{fig:setup_convergence}a). Sweeping
\(\varepsilon\) at resolution~4 against the exact spheroid resistances traces a
shallow U-curve minimised near \(\varepsilon\approx0.002\)--\(0.004\)
(Fig.~\ref{fig:setup_convergence}b), whose residual leading-order bias the
three-point Richardson extrapolation in \(\varepsilon\) removes for the
converged references of Section~\ref{sec:spectral_methods} (the fixed-%
\(\varepsilon\) dataset labels retain the sub-\(0.3\%\) bias quantified
above). This discretisation error is far below the percent-level truncation
errors examined in Section~\ref{sec:results}, so resolution~4 suffices.

\subsection{Spectral fidelity of the resistance labels}
\label{sec:spectral_methods}

The surrogate of Result~2 represents every particle by its
spherical-harmonic expansion truncated at degree \(L=15\), and is
trained on labels computed for exactly those band-limited surfaces.
Before reporting any surrogate accuracy we therefore answer a more
basic question: how much of a particle's hydrodynamic resistance is
lost by the representation itself? That is, when a general (not
band-limited) random particle is replaced by its degree-\(L\)
truncation, how large is the change in the resistance
matrix~\(\mathcal{R}\), and how does it depend on the particle's shape
and on~\(L\)? The answer doubles as a practical lookup rule for running
the solver on a new shape. The full spectral-convergence
analysis behind this subsection --- the per-mode Stokes sensitivity
kernels, the kernel-law collapse of the truncation error, and its
behaviour on anisotropic bases --- is given in the ESI, Section~S1.

\subsubsection{Protocol and error metric}

For every shape we build a triangulated surface on an icosphere of
resolution~\(r=4\) (\(V=2562\) vertices, \(F=5120\) faces) and
volume-normalise it to \(4\pi/3\), centred at the area-weighted
surface centroid (for the Gaussian-random-field (GRF) ensemble the
centring and scaling frame is frozen from the full-resolution
reference; for the CSH ensemble each truncation level is
independently normalised). The
regularised-Stokeslet--surface (RSS) solver \cite{Ferranti2024,Cortez2001,Cortez2005}
is run at three coarse regularisation amplitudes
\(\varepsilon\in\{0.005,\,0.0075,\,0.01\}\) and the resistance matrix
extrapolated to \(\varepsilon\to 0\) by the three-point quadratic
Richardson fit of Gallagher and Smith\cite{GallagherSmith2021}. This produces a
per-shape truth
\(\mathcal{R}_\infty \equiv \mathcal{R}(L\!=\!40,\,\varepsilon\!\to\!0)\).
The relative Frobenius error of a truncated GRM is reported per
block,
\[
  E_X(L) \;=\; \bigl\|X(\Gamma)-X(\Gamma_L)\bigr\|_F
                 \big/\bigl\|X(\Gamma)\bigr\|_F,
  \qquad X\in\{A,B,C\},
\]
with \(X(\Gamma_L)\) the corresponding block of the resistance
matrix solved on the surface rebuilt from SH coefficients truncated
at degree~\(L\). The \emph{required truncation degree} at
tolerance~\(\tau\) is defined monotonically,
\[
  L_{\rm required}^X(\tau)
   \;=\; \min\bigl\{L \,:\, E_X(L') < \tau \;\;\forall L'\ge L\bigr\},
\]
which guards against spurious crossings when \(E_X(L)\) momentarily
dips below the solver floor for very smooth shapes. We adopt
\(\tau=5\%\) as the engineering tolerance throughout.

The ensemble consists of \(N\approx2000\) random particles drawn from two
structurally different generators that between them span near-spherical
grains through very rough, strongly triaxial aggregates (RMS radial
roughness \(\varepsilon_{\rm rms}\) up to \({\sim}38\%\)).

The first family (\(N_{\rm GRF}\approx850\)) isolates roughness on an
otherwise spherical base. These are GRF radial perturbations of a unit
sphere,
\begin{equation}
  r(\theta,\varphi)=1+\sum_{\ell\geq 1}\sum_{m=-\ell}^{\ell}
     a_{\ell m}\,Y_\ell^m(\theta,\varphi),
  \label{eq:grf_shape}
\end{equation}
whose coefficients \(a_{\ell m}\) are independent zero-mean Gaussians with
a power-law variance spectrum, \(P_\ell=\sum_m a_{\ell m}^2\propto\ell^{-2\alpha}\);
we sweep the decay exponent from shallow to steep
(\(\alpha\in[0.2,\,2.7]\)) and the overall amplitude from smooth to rough
(RMS roughness \(\varepsilon_{\rm rms}\) up to \({\sim}30\%\)).

The second family (\(N_{\rm CSH}\approx1050\)) adds the anisotropic base a
sphere-baseline ensemble lacks. These are the Cartesian spherical-harmonic
(CSH) shapes of Section~\ref{sec:shapes}: a triaxial base ellipsoid
(degree \(\ell=1\), eqn~\eqref{eq:C1_block}) carrying the two-segment
roundness--roughness spectrum of
eqns~\eqref{eq:wei_two_segment}--\eqref{eq:random_Cl}. Because that
spectrum is band-limited at \(\ell=15\)---and a band-limited shape has
identically zero truncation error for \(L\geq 15\)---the spectrum is
continued for this study by a third power-law segment anchored to the
\(\ell=15\) amplitude,
\(d_\ell = q\,d_{15}\,(\ell/15)^{-\gamma}\) for \(16\leq\ell\leq 40\),
which models the unresolved sub-\(\mu\)CT roughness band identified by
surface-texture measurements\cite{Yang2016} in the fractal-continuation
spirit of Wei~et~al.\cite{WeiFractal2018} (\(q=0\) recovers the pure
band-limited spectrum). To populate the \((\alpha,\varepsilon_{\rm rms})\)
diagnostic plane broadly, the generator parameters are sampled more
widely than the training window of
eqn~\eqref{eq:wei_extended_ranges}: aspect ratio and flatness
\(\mathrm{AR},H\in[0.3,1]\), band energies up to
\(D_{2\text{--}8}\leq 1.5\) and \(D_{9\text{--}15}\leq 0.35\), slope
exponents \(\alpha\in[0.3,4.5]\) and \(\beta\in[1,2]\) (rather than the
natural-sand anchors), and continuation parameters \(q\in[0,1.5]\),
\(\gamma\in[0.5,2.5]\), reaching \(\varepsilon_{\rm rms}\) up to
\({\sim}38\%\).

\subsubsection{Two-number shape summary}
\label{sec:r1_decomposition}

The resistance blocks \(\boldsymbol{A}\) and \(\boldsymbol{C}\) are
symmetric rank-two tensors, and a symmetric \(3\times3\) tensor contains
only two irreducible pieces: a scalar trace (degree \(\ell=0\)) and a
symmetric-traceless part (degree \(\ell=2\))---the same decomposition the
surrogate's output heads use (Section~\ref{sec:model}). This constrains
which shape features can move the resistance at leading order. Take a
nearly spherical particle,
\(r(\theta,\varphi)=a\,[\,1+\varepsilon f(\theta,\varphi)\,]\) with
\(\varepsilon\ll1\) and \(f=\sum_{\ell m}f_{\ell m}Y_\ell^m\). ``First
order'' means expanding the resistance in the deformation amplitude
\(\varepsilon\) and keeping the term linear in \(\varepsilon\). That term
is a \emph{linear} functional of the shape perturbation \(f\). Since each
coefficient \(f_{\ell m}\) transforms under rotation as degree \(\ell\)
while the resistance correction transforms as a rank-two tensor
(\(\ell=0\oplus\ell=2\)), rotational symmetry---the Wigner--Eckart
selection rule---permits a nonzero linear coupling only when the shape
degree matches a degree present in the tensor: \(\ell=0\), which feeds
the trace, and \(\ell=2\), which feeds the traceless part. Every other
degree averages against the rank-two response to zero at first order and
so enters only at \(\mathcal{O}(\varepsilon^2)\). Brenner's classical
perturbation theory of a slightly deformed sphere\cite{Brenner1964a}
derives exactly this: to first order in the deformation amplitude only
the \(\ell=0\) and \(\ell=2\) surface harmonics enter the translational
and rotational resistance. The result has since been reformulated and
extended by Nourhani and Lammert\cite{NourhaniLammert2018} and by Nabil
and Nourhani\cite{NabilNourhani2024}.

Physically, the \(\ell=0\) mode rescales the mean radius (the particle's
size) and the \(\ell=2\) modes deform the sphere into a triaxial
ellipsoid, so the leading-order hydrodynamic content of any near-spherical
shape is precisely its size-and-ellipsoid fit. Volume normalisation
(Section~\ref{sec:canon}) fixes the \(\ell=0\) part and centring removes
the \(\ell=1\) part (a rigid shift of the centroid), leaving the base
ellipsoid at \(\ell=2\) and pure ``roughness'' at \(\ell\geq3\)---which,
by the selection rule, changes the resistance only at second order in its
amplitude.

This separation motivates summarising the roughness of any star-convex
particle by two numbers. The first is its total amplitude, the
above-base power
\[
  d_\infty^{\,2} \;\equiv\; \sum_{\ell\geq 3} P_\ell,
  \qquad
  P_\ell \;=\; \sum_m |c_{\ell m}|^2,
\]
whose physical meaning is the RMS radial roughness as a fraction of the
base radius, \(\varepsilon_{\rm rms}\equiv d_\infty/\sqrt{4\pi}\) (for the
CSH shapes, \(P_\ell\) sums the coefficient power over the three
Cartesian components). The second is how that power is distributed across
degrees: because natural roughness spectra are approximately power laws,
we capture the distribution by a single decay exponent \(\alpha\) defined
through \(P_\ell\propto\ell^{-2\alpha}\), fitted by log--log least squares
on \(\ell\in[3,20]\). The amplitude says how much roughness there is; the
exponent says how fine it is---and, as the next subsection shows, finer
(higher-\(\ell\)) roughness is hydrodynamically more expensive per unit
power, so both numbers are needed to predict the truncation error.

\subsubsection{Explaining convergence behaviour}

The truncation error is governed, to leading order, by the
\emph{kernel-weighted discarded power}
\begin{equation}
  E_X(L) \;\approx\; H_X(L) \;\equiv\;
     \sum_{\substack{\ell>L\\ \ell\geq 3}} K_X(\ell)\,P_\ell,
  \label{eq:kernel_tail}
\end{equation}
where \(K_X(\ell)\) is the per-mode Stokes sensitivity of block~\(X\),
measured directly from single-mode sphere perturbations (ESI,
Section~S1). Two measured properties of \(K_X\) govern everything
that follows: the kernel \emph{grows} with \(\ell\) --- finer surface
detail is hydrodynamically more expensive per unit power --- and the
ratio \(K_C/K_A\) decreases from \(4.94\) at \(\ell=2\) toward the
theoretical asymptote of \(3\) at large \(\ell\), so the rotation
block weighs its fine-scale tail roughly three times more heavily
than the translation block. Rotational resistance is therefore the
binding constraint on the truncation level throughout. The ESI
validates eqn~\eqref{eq:kernel_tail} directly on the ensemble
(log--log slope \(\approx 1\) on sphere-baseline shapes), quantifies
its breakdown on strongly anisotropic bases, and traces that
breakdown to an orientation-dependent linear response that no
rotation-invariant spectral summary can capture --- the information
the equivariant surrogate of Result~2 consumes.

\subsection{Irreducible structure of the grand resistance matrix}
\label{sec:irrep}

The grand resistance matrix $\boldsymbol{\mathcal{R}}$ of
eqn~\eqref{eq:grm} has three independent Cartesian blocks: the
symmetric translational resistance
$\boldsymbol{A}\in\mathbb{R}^{3\times 3}_{\mathrm{sym}}$, the symmetric
rotational resistance
$\boldsymbol{C}\in\mathbb{R}^{3\times 3}_{\mathrm{sym}}$, and the
general coupling $\boldsymbol{B}\in\mathbb{R}^{3\times 3}$. Under a
rigid rotation $\boldsymbol{R}\in\mathrm{SO}(3)$ of the particle each
block transforms as $\boldsymbol{X}\mapsto
\boldsymbol{R}\,\boldsymbol{X}\,\boldsymbol{R}^{\!\top}$. Each
symmetric $3\times 3$ matrix decomposes uniquely into the trivial
representation $\ell=0$ and the symmetric traceless $\ell=2$
representation,
\begin{equation}
\boldsymbol{A}
= \tfrac{1}{3}\operatorname{tr}(\boldsymbol{A})\,\boldsymbol{I}
  \;+\;\bigl[\boldsymbol{A}-\tfrac{1}{3}\operatorname{tr}(\boldsymbol{A})\,\boldsymbol{I}\bigr],
\label{eq:symm_decomp}
\end{equation}
and likewise for $\boldsymbol{C}$, yielding six independent components
each. The general $3\times 3$ tensor $\boldsymbol{B}$ further contains
an antisymmetric $\ell=1$ part, for a total of $6+6+9=21$ irreducible
degrees of freedom in $\boldsymbol{\mathcal{R}}$, organised as
$(2\!\times\!0^{e}\oplus 2\!\times\!2^{e})$ for the parity-even diagonal
blocks $\boldsymbol{A}$ and $\boldsymbol{C}$ and $(0^{o}\oplus
1^{o}\oplus 2^{o})$ for the parity-odd coupling block $\boldsymbol{B}$,
the superscripts $e$ and $o$ denoting even and odd parity\cite{Brenner1963,HappelBrenner1983,KimKarrila1991}.

The parity assignment is physical. Under the spatial inversion
$\boldsymbol{x}\mapsto -\boldsymbol{x}$ the diagonal blocks
$\boldsymbol{A}$ and $\boldsymbol{C}$ are invariant whereas
$\boldsymbol{B}$ changes sign, so any particle possessing a centre of
inversion has $\boldsymbol{B}=\boldsymbol{0}$ about that centre;
sufficiently rich point-group symmetry (axisymmetry, or three mutually
perpendicular mirror planes) likewise eliminates the block, whereas a
single mirror plane annihilates only part of
it\cite{HappelBrenner1983}. For the
statistically achiral microparticles that motivate this work the
relevant output is therefore
$2\!\times\!0^{e}\oplus 2\!\times\!2^{e}$, twelve components describing
$(\boldsymbol{A},\boldsymbol{C})$. The output therefore needs only parity-even heads. The \emph{hidden}
representation, however, is most accurate when it is mixed-parity: the
Cartesian spherical-harmonic input carries odd-parity (\(\ell=1^{o}\))
content, and retaining odd-parity hidden channels---which the trunk
recombines into the even-parity targets---improves accuracy over an
even-only hidden basis (ESI, Section~S5). SHEAR accordingly pairs a
mixed-parity hidden basis with parity-even output heads; an even-only
hidden basis would be smaller but structurally unable to produce a
nonzero \(\boldsymbol{B}\) and, as the ablation confirms, less accurate
on \((\boldsymbol{A},\boldsymbol{C})\).

\subsection{Equivariant neural surrogate}\label{sec:model}

The guiding requirement is simple to state: when the particle is
rotated, its predicted resistance tensors must rotate with it. We
want a network that obeys this automatically rather than approximately.
The training labels of Section~\ref{sec:solver} make the requirement
precise: under any rigid rotation
\(\boldsymbol{R}\in\mathrm{SO}(3)\) each degree-\(\ell\) coefficient
block transforms in both of its indices,
\(\boldsymbol{\mathsf{C}}_\ell\mapsto
\boldsymbol{R}\,\boldsymbol{\mathsf{C}}_\ell\,D^{(\ell)}(\boldsymbol{R})^{\!\top}\)
--- by \(\boldsymbol{R}\) in the Cartesian channel and by the Wigner
matrix \(D^{(\ell)}(\boldsymbol{R})\) in the harmonic index --- and the
labels rotate as
\(\boldsymbol{X}\mapsto\boldsymbol{R}\boldsymbol{X}\boldsymbol{R}^{\!\top}\)
(\(X\in\{\boldsymbol{A},\boldsymbol{C}\}\)). A regression model that
honours this transformation law by construction can never produce a
spurious orientation dependence; one that does not must learn it from
augmentation, at the cost of data efficiency and exact symmetry. We
build the former: the \emph{Spherical-Harmonic Equivariant Architecture
for Resistance} (SHEAR). Every operation in the network is an
\(\mathrm{SO}(3)\)-equivariant primitive from the \texttt{e3nn}
library\cite{Geiger2022}: Clebsch--Gordan-allowed tensor products
between irreducible representations, irrep-graded linear maps, and
parity-preserving gated nonlinearities. The composition of such primitives is itself equivariant. Within this architectural class the remaining
design choices---the irrep budget of the hidden layers, the maximum
coupling degree of the tensor products, and the form of the SPD output
head---are set by a systematic ablation over the design space, reported
in full in the Electronic Supplementary Information (Section~S5). A
defining principle of SHEAR is that it acts on a single, purely
equivariant pathway and carries no hand-computed invariant
side-channel (power spectrum or bispectrum). This is deliberate: the
power spectrum and bispectrum are themselves particular
Clebsch--Gordan contractions, so the equivariant trunk already
recovers this rotation-invariant content internally, and supplying it
again as an auxiliary input is redundant. The ablation of
Section~S5 bears this out, and SHEAR accordingly reaches lower error
with an order of magnitude fewer parameters than an
invariant-augmented variant.

\subsubsection{Input representation}
We supply the network with the full \(L_{\max}=15\) Cartesian
spherical-harmonic representation of Section~\ref{sec:shapes},
\(256\) coefficients per Cartesian channel. To break the multiplicative
ambiguity between the volume normalisation and the radial scale of
each shape we augment the three Cartesian channels with a fourth
channel carrying the spherical-harmonic expansion of the log radius
\(\log r(\theta,\varphi)\) evaluated on the canonicalised mesh, giving a
\((4\times 256)\) input tensor per shape. A Clebsch--Gordan transform
decomposes the low-degree part of this tensor on the fly into clean
\(\mathrm{SO}(3)\) irreps---explicit \(0^{e},1^{o},2^{e},\ldots\)
components---up to a tensor-product ceiling
\(\ell^{\mathrm{tp}}_{\max}=8\) (the vector channel contributing irreps
up to \(\ell=9\)). These irreps are the sole input to the equivariant
trunk. Spherical-harmonic degrees \(\ell>\ell^{\mathrm{tp}}_{\max}\)
would enter the model only through the optional invariant
side-channel that SHEAR omits, so its equivariant pathway consumes
shape information up to \(\ell\approx 8\). This is a deliberate
economy rather than a limitation: the resistance of a volume-normalised
grain is dominated by its low-degree, near-ellipsoidal structure, and
retaining the high-degree channel neither improves accuracy nor is
needed to reach it (Section~S5).

\subsubsection{Embedding and equivariant trunk}
\begin{figure*}[t]
\centering
\includegraphics[width=0.92\textwidth]{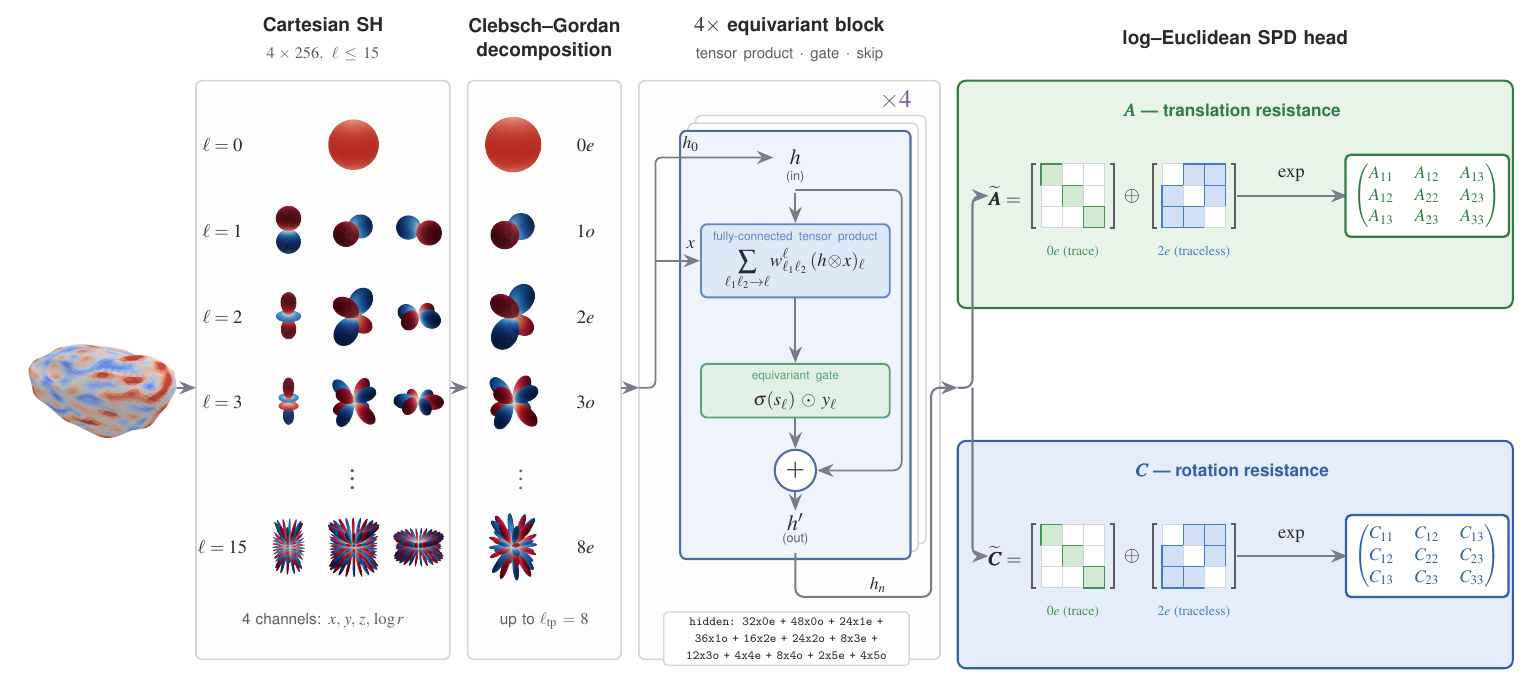}
\caption{\textbf{SHEAR architecture.}
  The \((4\times 256)\) Cartesian spherical-harmonic input is
  Clebsch--Gordan--decomposed into irreps (up to the tensor-product
  ceiling \(\ell^{\mathrm{tp}}_{\max}=8\)) and passed through a single
  equivariant trunk (four FCTP+Gate+linear-residual blocks); log-Euclidean
  \(\boldsymbol{A}\)- and \(\boldsymbol{C}\)-heads return
  \(1{\times}0^{e}\oplus1{\times}2^{e}\) per block, and the matrix
  exponential maps each to its SPD output. There is no invariant
  side-channel; every arrow is exactly \(\mathrm{SO}(3)\)-equivariant.}
\label{fig:shear_arch}
\end{figure*}

We collect the input irreps of the preceding paragraph into a single
feature \(\boldsymbol{x}=\bigoplus_{c}\boldsymbol{x}^{(\ell_c,p_c)}\),
a direct sum of vectors each carrying a definite degree \(\ell_c\) and
parity \(p_c\in\{e,o\}\) and transforming under a rotation
\(\boldsymbol{R}\) by the corresponding Wigner matrix
\(D^{(\ell_c)}(\boldsymbol{R})\). An equivariant linear map
\(\boldsymbol{W}_{\!\mathrm{emb}}\), which is block-diagonal in \((\ell,p)\)
and acts only within channels of matching degree and parity, embeds
\(\boldsymbol{x}\) into a hidden multi-parity budget
\begin{equation}
\begin{aligned}
\mathcal{H} = {}&32{\times}0^{e}\oplus 48{\times}0^{o}
              \oplus 24{\times}1^{e}\oplus 36{\times}1^{o}\\
            &\oplus 16{\times}2^{e}\oplus 24{\times}2^{o}
              \oplus 8{\times}3^{e}\oplus 12{\times}3^{o}\\
            &\oplus 4{\times}4^{e}\oplus 8{\times}4^{o}
              \oplus 2{\times}5^{e}\oplus 4{\times}5^{o},
\end{aligned}
\label{eq:hidden_irreps}
\end{equation}
so that \(\boldsymbol{h}_0=\boldsymbol{W}_{\!\mathrm{emb}}\boldsymbol{x}\).
The trunk then applies four residual blocks
\begin{equation}
\boldsymbol{h}_{k}
\;=\;
\boldsymbol{W}_{k}\,\boldsymbol{h}_{k-1}
\;+\;\mathcal{G}\bigl[\,\mathcal{T}(\boldsymbol{h}_{k-1},\boldsymbol{x})\,\bigr],
\qquad k=1,\ldots,4,
\label{eq:equivariant_block}
\end{equation}
where \(\boldsymbol{W}_{k}\) is an equivariant linear (skip) map,
\(\mathcal{T}\) is a tensor product that couples the running features to
the input, and \(\mathcal{G}\) is a gated nonlinearity
(Fig.~\ref{fig:shear_arch}).

\subsubsection{Tensor product and gate}
The tensor product is where angular information is mixed. Two irreps of
degrees \(\ell_1,\ell_2\) combine, through the Clebsch--Gordan
coefficients \(C^{\ell\,m}_{\ell_1 m_1,\,\ell_2 m_2}\), into every output
degree \(\ell\) that the selection rule
\(|\ell_1-\ell_2|\le\ell\le\ell_1+\ell_2\) permits (with parity
\(p_1p_2\)):
\begin{equation}
\bigl(\boldsymbol{u}^{(\ell_1)}\!\otimes_{w}\boldsymbol{v}^{(\ell_2)}\bigr)^{(\ell)}_{m}
=w_{\ell_1\ell_2\ell}\!\!\sum_{m_1,m_2}\!
   C^{\ell\,m}_{\ell_1 m_1,\,\ell_2 m_2}\,
   u^{(\ell_1)}_{m_1}\,v^{(\ell_2)}_{m_2}.
\label{eq:tensor_product}
\end{equation}
\(\mathcal{T}\) applies eqn~\eqref{eq:tensor_product} over all admissible
degree pairs up to a coupling ceiling \(\ell^{\mathrm{tp}}_{\max}=8\),
with one learnable weight \(w_{\ell_1\ell_2\ell}\) per allowed path
(``fully connected'' over the channel multiplicities). Because the
Clebsch--Gordan map is itself equivariant, so is every such product.
The gate \(\mathcal{G}\)\cite{Weiler2018} then supplies the
nonlinearity without breaking equivariance: each non-scalar output irrep
\(\boldsymbol{y}^{(\ell)}\) of \(\mathcal{T}\) is rescaled by a
sigmoid-activated \emph{invariant} gate scalar \(s_\ell\) taken from the
block's own \(0^{e}\) channels,
\begin{equation}
\mathcal{G}:\;
\boldsymbol{y}^{(\ell)}\;\longmapsto\;\sigma(s_\ell)\,\boldsymbol{y}^{(\ell)}
\quad(\ell\ge 1),
\label{eq:gate}
\end{equation}
while the scalar channels pass through a pointwise nonlinearity
(SiLU on the even-parity scalars; tanh, an odd function, on the
odd-parity pseudoscalars, which preserves their sign flip under
inversion).
Multiplying an irrep by a rotation-invariant scalar preserves its
transformation law, so eqn~\eqref{eq:gate} is equivariant. Crucially, the
gate scalars are computed \emph{inside} the trunk from the running
features and are never injected from outside, which is what keeps the
whole network on a single equivariant pathway.

\subsubsection{Symmetric-positive-definite output heads}
Each block is recovered from the final features \(\boldsymbol{h}_4\)
through a physically constrained head. A symmetric \(3\times3\) tensor
decomposes irreducibly into a trace part (\(0^{e}\)) and a
traceless-symmetric part (\(2^{e}\)), so an equivariant linear map
\(\boldsymbol{W}^{X}\!:\mathcal{H}\to 1{\times}0^{e}\oplus1{\times}2^{e}\)
followed by the standard embedding \(\iota(\cdot)\) of these six
coefficients into a symmetric matrix yields a symmetric
\(\widetilde{\boldsymbol{X}}=\iota(\boldsymbol{W}^{X}\boldsymbol{h}_4)\)
for each \(X\in\{\boldsymbol{A},\boldsymbol{C}\}\). We interpret
\(\widetilde{\boldsymbol{X}}\) as the matrix logarithm of the resistance
block and recover the prediction by the symmetric matrix exponential.
Writing the eigendecomposition
\(\widetilde{\boldsymbol{X}}=\boldsymbol{Q}\,\boldsymbol{\Lambda}\,\boldsymbol{Q}^{\!\top}\)
with \(\boldsymbol{Q}\in\mathrm{O}(3)\) and
\(\boldsymbol{\Lambda}=\operatorname{diag}(\lambda_1,\lambda_2,\lambda_3)\),
\begin{equation}
\boldsymbol{X}
=\exp\!\bigl(\widetilde{\boldsymbol{X}}\bigr)
=\boldsymbol{Q}\,\operatorname{diag}\!\bigl(e^{\lambda_1},e^{\lambda_2},e^{\lambda_3}\bigr)\,\boldsymbol{Q}^{\!\top},
\label{eq:spd_head}
\end{equation}
which is symmetric positive-definite for \emph{any} symmetric
\(\widetilde{\boldsymbol{X}}\) (the eigenvalues \(e^{\lambda_i}>0\)), so
the head cannot emit a non-physical resistance in exact arithmetic (in
float32 the exponential can run away for extreme out-of-distribution
inputs; quantified in the ESI, Section~S5). The map is also exactly
equivariant: under a rotation
\(\widetilde{\boldsymbol{X}}\mapsto\boldsymbol{R}\widetilde{\boldsymbol{X}}\boldsymbol{R}^{\!\top}\)
the eigenvectors rotate to \(\boldsymbol{R}\boldsymbol{Q}\) while the
eigenvalues are unchanged, giving
\(\exp(\boldsymbol{R}\widetilde{\boldsymbol{X}}\boldsymbol{R}^{\!\top})
=\boldsymbol{R}\,\boldsymbol{X}\,\boldsymbol{R}^{\!\top}\). Training
targets are the corresponding matrix logarithms
\(\widetilde{\boldsymbol{X}}=\log\boldsymbol{X}
=\boldsymbol{Q}\operatorname{diag}(\log\lambda_i)\boldsymbol{Q}^{\!\top}\),
which are well defined because the solver labels are SPD
(\(\lambda_i>0\)). Regressing in these log coordinates is what makes the
positive-definiteness constraint and the wide dynamic range of the
targets tractable. The matrix logarithm maps the curved cone of SPD
matrices onto the flat vector space of all symmetric matrices, where
ordinary (Euclidean) regression is well posed and, through the
exponential of eqn~\eqref{eq:spd_head}, cannot leave the cone. The
distance this induces back on the SPD matrices is the log-Euclidean
Riemannian metric of Arsigny \emph{et al.}\cite{Arsigny2006}. Because the eigenvalues of \(\boldsymbol{A}\) and
\(\boldsymbol{C}\) span more than a decade, taking their logarithm
compresses the regression targets to an \(\mathcal{O}(1)\) range and
places the overall size of each tensor (the trace of the logarithm) and
its anisotropy (the traceless part) on an equal footing.

\subsubsection{Training, loss, and augmentation}
We minimise a loss that combines a log-Euclidean term, matching the
predicted and target matrix logarithms, with a relative-Frobenius term
on the recovered SPD matrices,
\begin{equation}
\mathcal{L}
= \!\!\sum_{X\in\{A,C\}}\!\!
  \Bigl[
    \bigl\lVert\log X-\log X^{\mathrm{pred}}\bigr\rVert_{F}^{2}
    + \lambda\,
    \frac{\lVert X-X^{\mathrm{pred}}\rVert_{F}}{\lVert X\rVert_{F}}
  \Bigr],
\label{eq:loss}
\end{equation}
with \(\lambda=0.3\), optimised by AdamW\cite{Loshchilov2019} at base
learning rate \(3\!\times\!10^{-4}\), weight decay \(10^{-5}\), batch
size \(1024\) and a cosine schedule to \(10^{-5}\) over \(100\) epochs.
Crucially, the network requires \emph{no} data augmentation. Because
every layer is equivariant by construction, the shape-to-resistance map
already transforms correctly under every rotation and reflection, so the
rotation- and mirror-augmentation that non-equivariant models depend on
to learn orientation is unnecessary. Each shape is seen in a single
orientation during training. The ablation of Section~S5 confirms this
directly---adding such augmentation changes the sealed-test error only
within run-to-run noise. The model is trained on
\(110\,000\) shapes with \(10\,000\) held out for validation. All
results below are reported on a sealed test set of \(18\,123\) shapes
never seen during training or model selection, and the reported
checkpoint is the one minimising the validation \(\boldsymbol{C}\)
error.

\subsection{Settling, rotation, and Brownian-dynamics observables}
\label{sec:settling_methods}

Each of the three downstream regimes we examine — gravitational
settling, externally driven rotation, and Brownian diffusion — follows
from a closed-form evaluation of the resistance blocks
\(\boldsymbol{A}\) and \(\boldsymbol{C}\) predicted by the surrogate,
with no additional solver calls. The observables are defined as
follows.

\subsubsection{Gravitational settling (translation block \(\boldsymbol{A}\))}
The terminal-velocity vector is
\(\boldsymbol{U}=\boldsymbol{A}^{-1}\boldsymbol{F}_{\!g}/\mu\), where
\(\boldsymbol{F}_{\!g}=(\rho_p-\rho_f)V_p\,\boldsymbol{g}\) is the
buoyancy-corrected weight of the particle.  When
\(\boldsymbol{A}\) has distinct eigenvalues, \(\boldsymbol{U}\) is in
general not parallel to gravity. We characterise the resulting
deviation by the drift angle
\(\theta=\angle(\boldsymbol{U},\boldsymbol{g})\) and the per-shape
speed anisotropy \(|\boldsymbol{U}|_{\max}/|\boldsymbol{U}|_{\min}\)
over 64 uniformly random body orientations of each of the
\(18\,123\) sealed test shapes (\(1.16\!\times\!10^{6}\) settling
events in total). These are instantaneous, fixed-orientation drift statistics rather than
full free-settling trajectories: an asymmetric grain also reorients as
it settles, driven by the gravitational torque that arises when its
centre of mass does not coincide with its centre of resistance. This
reorientation is an achiral effect --- it requires no chiral shape ---
and lies outside the fixed-orientation,
$(\boldsymbol{A},\boldsymbol{C})$-only treatment reported here. The Happel--Brenner orientation-averaged scalar friction
is\cite{HappelBrenner1983,Brenner1963}
\(K_{\rm avg}\equiv\tfrac{1}{3}\,\mathrm{tr}(\boldsymbol{A})\).

\subsubsection{Externally driven rotation (rotation block \(\boldsymbol{C}\))}
An external torque drives an angular velocity
\(\boldsymbol{\Omega}=\mu^{-1}\boldsymbol{C}^{-1}\boldsymbol{T}_{\rm ext}\).
We sample 64 uniformly random unit torque directions per shape and
measure the misalignment angle
\(\Phi=\angle(\boldsymbol{\Omega},\boldsymbol{T}_{\rm ext})\) and
the rotational speed anisotropy
\(|\boldsymbol{\Omega}|_{\max}/|\boldsymbol{\Omega}|_{\min}\). A
unit external torque is the most direct proxy for any mechanism that
imposes a hydrodynamic torque — magnetic-bead or
optical-trap microrheology, and the centre-of-mass-offset gravity
torque on asymmetric sedimenters. Coupling the body to a background
shear additionally requires the strain-coupling block of the
\(11\times 11\) Faxén-extended grand mobility, which the \(6\times 6\) GRM
does not predict. The orientation-averaged rotational scalar is
\(K_{\rm avg}^{C}\equiv\tfrac{1}{3}\,\mathrm{tr}(\boldsymbol{C})\).

\subsubsection{Brownian diffusion (\(\boldsymbol{A}^{-1}\) and
\(\boldsymbol{C}^{-1}\))}
The orientation-averaged translational diffusion coefficient is
\(D_T=(k_B T/3\mu)\,\mathrm{tr}(\boldsymbol{A}^{-1})\); the
rotational analogue is
\(D_R=(k_B T/3\mu)\,\mathrm{tr}(\boldsymbol{C}^{-1})\). The classical
baselines are the volume-equivalent Stokes--Einstein sphere
\cite{HappelBrenner1983}, the Perrin spheroid \cite{Perrin1934}, and
the Bagheri \& Bonadonna empirical Stokes-limit drag
\cite{Bagheri2016}.

\subsubsection{Stokes-limit form of the empirical correlations}
The drag correlations we benchmark against were calibrated to the
\emph{finite}-Reynolds-number drag of freely falling
particles\cite{Bagheri2016,Holzer2008,Ganser1993}. To place them on the
same footing as the zero-Reynolds-number resistance matrix we reduce
each to its Stokes (\(\mathrm{Re}\!\to\!0\)) limit, where it predicts a
single orientation-averaged translational friction \(\zeta_t\)
(Table~\ref{tab:stokes_limit}), reducing to the Stokes sphere
\(\zeta_t=6\pi\mu R_{\rm eq}\) when the shape is spherical. As these are
experimental fits for the orientation-averaged scalar drag, they carry
no tensor orientation. We therefore benchmark them only against the
orientation-averaged scalar \(K_{\rm avg}=\tfrac13
\mathrm{tr}(\boldsymbol{A})\) of the Stokes ground truth (comparing
\(\zeta_t/\mu\) with \(K_{\rm avg}\); the relative errors are
\(\mu\)-independent), the comparison most favourable to the
correlations.

\begin{table}[!htbp]
\centering
\small
\caption{Stokes-limit (\(\mathrm{Re}\!\to\!0\)) reduction of the
  benchmarked drag correlations to an orientation-averaged
  translational friction \(\zeta_t\). Here \(d_{\rm eq}\) is the
  volume-equivalent diameter, \(\psi\) and \(\psi_\parallel\) the
  overall and lengthwise sphericities, \(d_n\) is Ganser's
  projected-area-equivalent diameter --- evaluated here as
  \(d_n=d_{\rm eq}\), exact for the sphere and a standard
  approximation for near-isometric grains --- and the Bagheri form
  factor is
  \(k_S=\tfrac12(F_S^{1/3}+F_S^{-1/3})\) with
  \(F_S=f\,e^{1.3}\,d_{\rm eq}^{3}/(LIS)\), flatness \(f=S/I\),
  elongation \(e=I/L\), and principal-axis lengths
  \(L\!\ge\!I\!\ge\!S\). Mesh-based definitions of every descriptor
  are given in the ESI, Section~S4. Every row reduces to
  \(\zeta_t=3\pi\mu d_{\rm eq}=6\pi\mu R_{\rm eq}\) for a sphere.}
\label{tab:stokes_limit}
\setlength{\tabcolsep}{5pt}
\begin{tabular}{@{}ll@{}}
\hline\hline
Correlation & Stokes-limit friction \(\zeta_t\) \\
\hline
Ganser 1993 & \(3\pi\mu d_{\rm eq}\bigl[\tfrac13\,(d_n/d_{\rm eq})+\tfrac23\psi^{-1/2}\bigr]\) \\[2pt]
H\"olzer \& Sommerfeld 2008 & \(\pi\mu d_{\rm eq}\bigl(\psi_\parallel^{-1/2}+2\psi^{-1/2}\bigr)\) \\[2pt]
Bagheri \& Bonadonna 2016 & \(3\pi\mu d_{\rm eq}\,k_S\) \\
\hline\hline
\end{tabular}
\end{table}

\section{Results}\label{sec:results}

The three results follow in turn: the fidelity of the \(\ell\le15\) representation, the accuracy and speed of the surrogate trained on it, and the transport physics that the predicted tensors reveal.

\subsection{Spectral fidelity of the resistance labels}
\label{sec:spectral_results}

\begin{figure*}[t]
  \centering
  \includegraphics[width=0.92\textwidth]{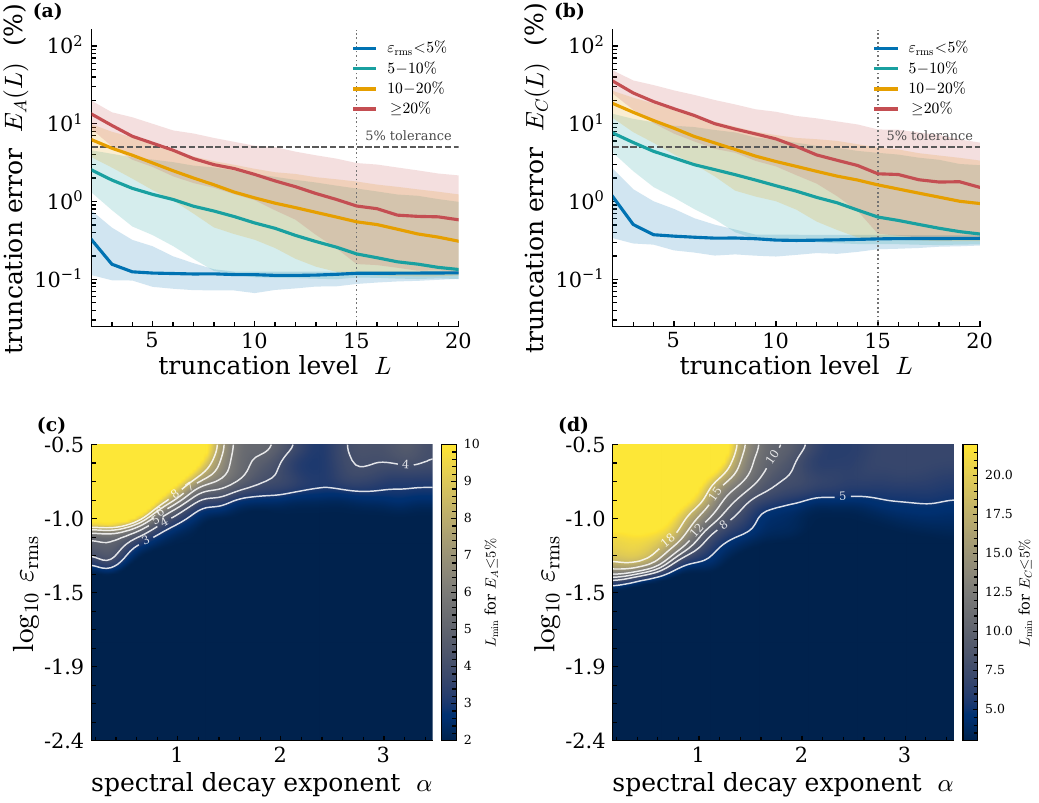}
  \caption{\textbf{Spectral fidelity of the resistance labels on the
    pooled random-particle ensemble} (\(N\approx2000\): \({\sim}850\)
    sphere-baseline GRF + \({\sim}1050\) triaxial-base CSH shapes).
    \emph{(a, b)} Per-band median (line) and inter-quartile band
    (shading) of the truncation error \(E_A(L)\) and \(E_C(L)\) in
    percent, stratified into four fixed physical-amplitude bands of
    \(\varepsilon_{\rm rms}\); the horizontal dashed line marks the
    \(5\%\) engineering tolerance and the vertical dotted line the
    dataset truncation \(L=15\). At \(L=15\) the median errors are
    \(0.14\%\) \((A)\) and \(0.38\%\) \((C)\), the 90th percentiles
    \(2.8\%\) and \(7.9\%\), and \(94\%\)/\(81\%\) of shapes are within
    tolerance (cumulative distributions in the ESI, Section~S1).
    \emph{(c, d)} Smoothed maps of \(L_{\rm required}^{A}(5\%)\) and
    \(L_{\rm required}^{C}(5\%)\) over
    \((\alpha,\,\log_{10}\varepsilon_{\rm rms})\); white contours
    mark integer levels. Note the different colour scales: the
    \(A\)-block map saturates near \(L\!\approx\!10\) while the
    \(C\)-block extends to \(L\!\approx\!22\) --- rotation is the
    binding block.}
  \label{fig:result1_main}
\end{figure*}

Figure~\ref{fig:result1_main} summarises the truncation behaviour
over the pooled ensemble of \({\approx}2000\) random particles.

Figs.~\ref{fig:result1_main}a and~\ref{fig:result1_main}b trace how
the truncation error falls with the SH degree.
Each curve is the median \(E_X(L)\) for shapes binned by their RMS
roughness amplitude, with the inter-quartile band shaded. The bins
are fixed in physical units --- \(\varepsilon_{\rm rms}<5\%\)
(smooth), \(5\)--\(10\%\), \(10\)--\(20\%\), and \(\geq 20\%\) (very
rough). Smoother shapes converge faster, rougher shapes need more SH
degrees, and the gap between translation and rotation is pronounced:
at the \(5\%\) tolerance (dashed line), smooth shapes typically cross
by \(L\approx 3\)--\(5\) for \(A\) but \(L\approx 6\)--\(10\) for
\(C\), and the rough band needs \(L\approx 8\)--\(15\) for \(A\)
versus \(L\approx 15\)--\(20\) for \(C\). This factor of two-to-three
asymmetry between the blocks is exactly the measured
\(K_C/K_A\) kernel weighting of eqn~\eqref{eq:kernel_tail} at work
(ESI, Section~S1).

These same data answer the question this study needs: what fraction of
random particles is fully converged at a given truncation? At
\(L=15\) --- the representation used for the Result-2 dataset --- the
translation block is within the \(5\%\) tolerance for \(94\%\) of
shapes and the rotation block for \(81\%\) (the full cumulative
distributions over \(L\) are given in the ESI, Section~S1), with
median errors of only \(0.14\%\) \((A)\) and \(0.38\%\) \((C)\) and
90th percentiles of \(2.8\%\) and \(7.9\%\).
Medians and counts emphasise different aspects of the same
heavy-tailed error distributions: the median curves of
Figs.~\ref{fig:result1_main}a and~\ref{fig:result1_main}b lie far below
tolerance at \(L=15\), while the \(19\%\) of
rotation-block failures occur in the upper tail of
each band, with a pass rate that falls monotonically with roughness
--- from \(98\%\) in the smoothest band to \(78\%\), \(74\%\) and
\(65\%\) across the three rougher bands.
Two consequences matter for the rest of this study. First, the
Result-2 training shapes are band-limited at \(\ell\leq 15\) by
construction, so their solver labels carry \emph{no} truncation error.
Second, when the trained surrogate is applied to a natural
particle via its \(\ell\leq 15\) SH fit, the representation itself
costs a median of only \(\lesssim 0.4\%\) in either block --- far smaller
than the surrogate's own prediction error (Result~2), so the \(\ell\leq15\)
truncation is not the accuracy-limiting step --- and the phase map
below flags the rare rough outliers for which a higher-fidelity
treatment is warranted.

Figs.~\ref{fig:result1_main}c and~\ref{fig:result1_main}d turn the
convergence behaviour into an operational lookup table. The heatmaps
of \(L_{\rm required}^A(5\%)\) and \(L_{\rm required}^C(5\%)\) over the
diagnostic plane
\((\alpha,\,\log_{10}\varepsilon_{\rm rms})\) --- smoothed by
\(k\)-NN median regression --- organise along monotone contours:
\(L_{\rm required}\) decreases with increasing \(\alpha\) (steeper
spectral decay) and with decreasing \(\varepsilon_{\rm rms}\) (less
above-base roughness). The \(A\)-block map saturates near
\(L\approx 10\) even in the roughest corner, while the \(C\)-block
scale extends to \(L\approx 22\): rotation is the binding
constraint, in line with the measured \(K_C/K_A\) weighting. For
practical use, a particle parameterised by
\((\alpha,\,\varepsilon_{\rm rms})\) maps to a one-glance
recommendation: read the contour level closest to its coordinates on
the binding (\(C\)) map and that is the \(L\) at which the full
resistance matrix will be inside the chosen tolerance.

\subsection{Surrogate accuracy}\label{sec:surrogate_results}

Result~1 established that the $\ell\!\leq\!15$ spherical-harmonic representation and its solver labels are faithful to well within the surrogate's own error; this section quantifies how accurately the equivariant network reproduces those labels.

\begin{figure*}[t]
\centering
\includegraphics[width=\textwidth]{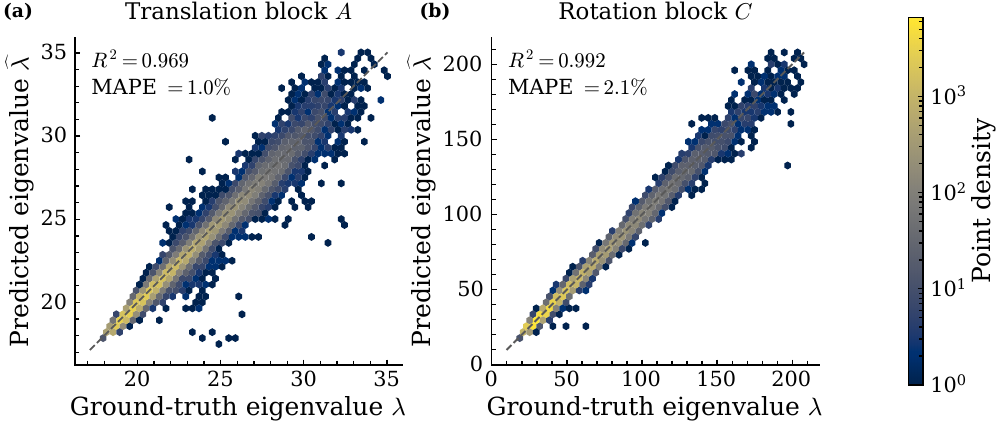}
\caption{\textbf{SHEAR accuracy on the sealed test set.}
  \textbf{(a,b)}~Predicted versus ground-truth eigenvalues of the translation
  block \(\boldsymbol{A}\) and rotation block \(\boldsymbol{C}\) across the
  \(18{,}123\)-shape sealed test set (density on a logarithmic scale, identity
  line dashed), at a mean absolute percentage error of \(1.0\,\%\)
  (\(\boldsymbol{A}\)) and \(2.1\,\%\) (\(\boldsymbol{C}\)).}
\label{fig:shear}
\end{figure*}

\begin{figure*}[t]
  \centering
  \includegraphics[width=0.92\textwidth]{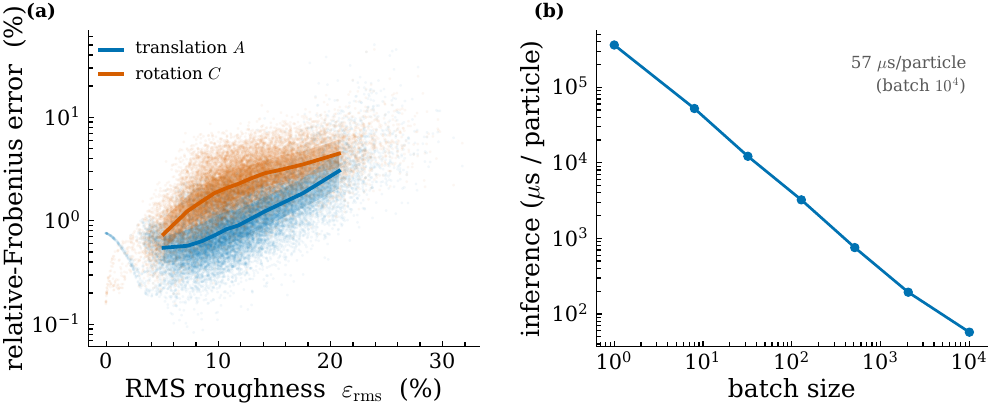}
  \caption{\textbf{In-distribution accuracy and speed of SHEAR.}
    \textbf{(a)}~Relative-Frobenius error of \(\boldsymbol{A}\) and
    \(\boldsymbol{C}\) versus the RMS roughness
    \(\varepsilon_{\rm rms}\): faint points are individual test shapes,
    the solid lines are the binned median, and the shaded ribbons the
    inter-quartile range. The error grows smoothly with roughness and
    the rotational block is uniformly harder than the translational
    one.
    \textbf{(b)}~Batched inference throughput on a single commodity GPU,
    reaching \(57\,\mu\mathrm{s}\) per shape at batch \(10^{4}\).}
  \label{fig:shear_blocks}
\end{figure*}

Across the sealed test set SHEAR predicts the full translational and
rotational resistance blocks at a mean relative-Frobenius error of
\(1.4\,\%\) on \(\boldsymbol{A}\) (median \(1.0\,\%\)) and \(2.7\,\%\)
on \(\boldsymbol{C}\) (median \(2.3\,\%\)). The predicted eigenvalue
spectra lie on the identity line across over an order of magnitude
(Fig.~\ref{fig:shear}a,b). The error grows smoothly with roughness
(Fig.~\ref{fig:shear_blocks}a; Table~\ref{tab:r2_accuracy}): the median
relative-Frobenius error rises monotonically with the RMS roughness
\(\varepsilon_{\rm rms}\), from \(\approx0.7\,\%\) (\(\boldsymbol{A}\))
and \(1.0\,\%\) (\(\boldsymbol{C}\)) for the smoothest, near-spherical
grains to \(\approx3\,\%\) and \(4\,\%\) for the roughest, most
aspherical aggregates. The rotational block is
uniformly harder than the translational one, as expected:
\(\boldsymbol{C}\) scales with the cube of the body size (longest axis
for tumbling) and
is the more sensitive to the high-degree surface roughness that an
\(L_{\max}=15\) representation resolves least well.

\begin{table}[!htbp]
\centering
\caption{Mean relative-Frobenius error of SHEAR on the
  \(18\,123\)-shape sealed test set, stratified into three bands of the
  high-degree roughness \(D_{9{-}15}\) of Section~\ref{sec:shapes}
  (smooth ellipsoids, \(D_{9{-}15}{=}0\), fall in the first band).
  Float32 inference.}
\label{tab:r2_accuracy}
\begin{tabular}{lccc}
\hline\hline
Roughness band \(D_{9{-}15}\) & \(n\) & \(\boldsymbol{A}\) (\%) & \(\boldsymbol{C}\) (\%) \\
\hline
\([0,\,0.05)\)        & 6\,937  & 0.8 & 1.8 \\
\([0.05,\,0.10)\)     & 6\,273  & 1.6 & 2.8 \\
\([0.10,\,0.15]\)     & 4\,913  & 1.9 & 3.7 \\
\hline
\textbf{all}             & \textbf{18\,123} & \textbf{1.4} & \textbf{2.7} \\
\hline\hline
\end{tabular}
\end{table}

Equivariance itself is not learned but built in: because every layer is an
\(\mathrm{SO}(3)\)-equivariant primitive, the conjugation law
\(\boldsymbol{X}(D(\boldsymbol{R})\boldsymbol{x})=\boldsymbol{R}\,\boldsymbol{X}(\boldsymbol{x})\,\boldsymbol{R}^{\!\top}\)
holds by construction, and a direct check (\(96\) shapes \(\times\)
\(256\) rotations) confirms it at the level of accumulated float32
round-off (median residual \(1\times10^{-6}\)).

To isolate what equivariance contributes, a non-equivariant control of
matched size is trained: a four-layer multilayer perceptron
(\(0.92\)M parameters, comparable to SHEAR's \(0.98\)M) on
rotation-invariant descriptors of the shape (the \(\ell\le10\) power
spectrum and cross-channel bispectrum, \(247\) features), regressing the
six eigenvalues of \(\boldsymbol{A}\) and \(\boldsymbol{C}\). This
baseline is specialised to exactly the rotation-invariant magnitudes,
yet SHEAR matches or exceeds it even there---a mean absolute percentage
error (MAPE, the mean of \(|\hat{\lambda}-\lambda|/|\lambda|\) over the
eigenvalues) of \(1.0\,\%\)/\(2.1\,\%\)
(\(\boldsymbol{A}\)/\(\boldsymbol{C}\)) for SHEAR against
\(1.0\,\%\)/\(3.2\,\%\) for the baseline.
The decisive difference, however, is orientation: a model with
rotation-invariant inputs is necessarily blind to the eigenvectors and
can place no tensor in the laboratory frame. It therefore cannot
recover the oriented, traceless part of \(\boldsymbol{A}\) and
\(\boldsymbol{C}\)---the content responsible for every
orientation-dependent observable of
Section~\ref{sec:settling_results} (lateral drift, rotational
misalignment, diffusion-tensor anisotropy)---returning there the same
null answer as the classical scalar drag correlations. Equivariance is
what turns an accurate spectrum into a usable tensor, and it does so
here at \emph{no} cost in spectral accuracy.

Speed compounds these advantages. At batch \(10^4\) SHEAR evaluates the full grand resistance tensors at
\(57\,\mu\mathrm{s}\) per shape (\(1.7\times10^{4}\) shapes\,s\(^{-1}\))
on a single commodity GPU (Fig.~\ref{fig:shear_blocks}b), against the
several seconds of a regularised-Stokeslet solve---a
speed-up of four to five orders of magnitude. Exact equivariance
compounds this: a single forward pass fixes the tensor for \emph{all}
orientations of a shape, since any rotated configuration is recovered by
a \(3\times3\) conjugation at no model cost, so an orientation-resolved
settling or Brownian-dynamics ensemble pays the network cost only once
per distinct morphology.

\subsection{Settling, rotation, and Brownian consequences}
\label{sec:settling_results}

An accurate surrogate matters only if its tensor output changes the predicted physics. This section therefore closes by tracing the settling, rotational, and Brownian consequences that the full anisotropic $\boldsymbol{A}$ and $\boldsymbol{C}$ capture and that every scalar or spheroid reduction misses.

The atmosphere holds roughly four times more coarse dust than every
operational climate model simulates, a discrepancy that
translates\cite{Adebiyi2020} to an additional radiative forcing of
order \(+0.15\,\mathrm{W\,m}^{-2}\).  Individual Saharan mineral grains up to
\(\sim 450\,\mu\mathrm{m}\) routinely survive 2400--3500 km of
transatlantic transport that no sphere-equivalent settling
parameterisation predicts \cite{vanderDoes2018}, and matching in-situ
Cabo Verde observations requires an artificial \(40\,\%\) reduction in
the modelled settling velocity \cite{Drakaki2022}. Microplastic fibres
are now recovered from every region of human lung tissue
\cite{Jenner2022} and settle up to \(76\,\%\) slower than
equivalent-volume spheres \cite{Tatsii2024}, while the river-to-ocean
flux estimate carries 2--3 orders of magnitude of uncertainty
\cite{Weiss2021,vanSebille2020}. These models share one structural
defect: each compresses \(\boldsymbol{A}\) to a single scalar, so the
predicted terminal velocity is parallel to gravity for every particle,
with no orientation dependence and no shape-induced variance.

\begin{figure*}[t!]
  \centering
  \begin{subfigure}[t]{0.31\textwidth}
    \centering
    \includegraphics[width=\linewidth]{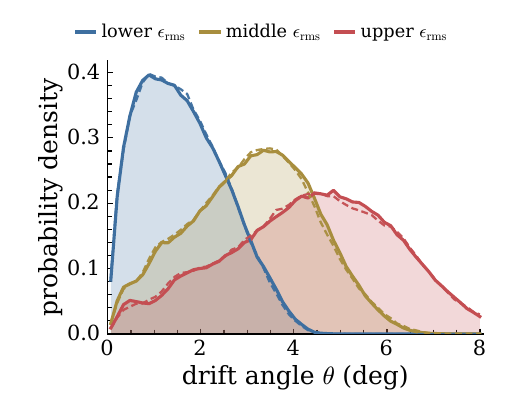}
    \caption{Settling drift angle.}
    \label{fig:result3_settling_drift}
  \end{subfigure}\hfill
  \begin{subfigure}[t]{0.31\textwidth}
    \centering
    \includegraphics[width=\linewidth]{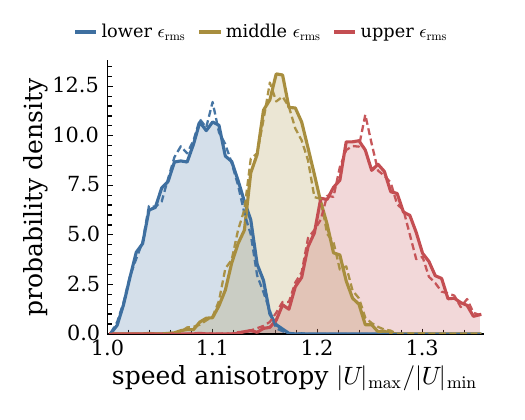}
    \caption{Settling speed anisotropy.}
    \label{fig:result3_settling_speed}
  \end{subfigure}\hfill
  \begin{subfigure}[t]{0.31\textwidth}
    \centering
    \includegraphics[width=\linewidth]{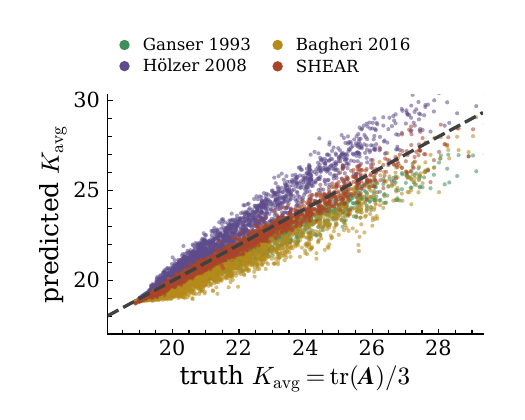}
    \caption{Scalar friction \(K_{\rm avg}\), predicted vs measured.}
    \label{fig:result3_settling_friction}
  \end{subfigure}\\[4pt]
  \begin{subfigure}[t]{0.31\textwidth}
    \centering
    \includegraphics[width=\linewidth]{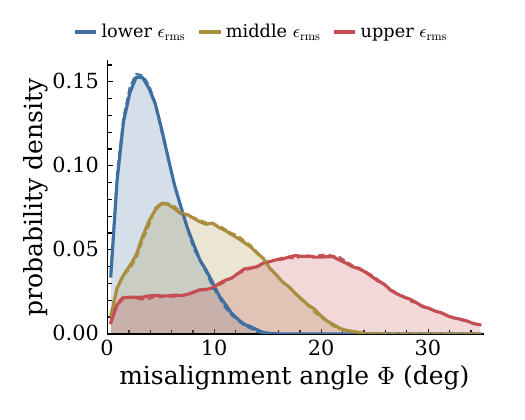}
    \caption{Rotational misalignment angle.}
    \label{fig:result3_rotation_misalign}
  \end{subfigure}\hfill
  \begin{subfigure}[t]{0.31\textwidth}
    \centering
    \includegraphics[width=\linewidth]{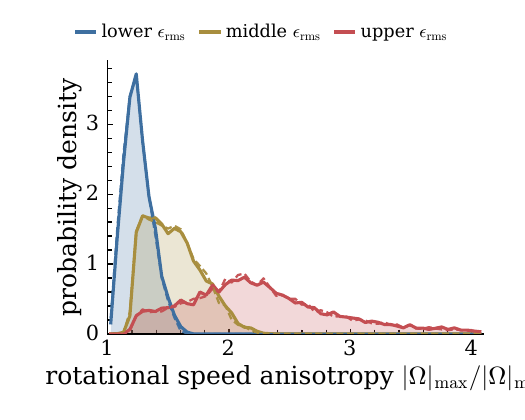}
    \caption{Rotational speed anisotropy.}
    \label{fig:result3_rotation_anisotropy}
  \end{subfigure}\hfill
  \begin{subfigure}[t]{0.31\textwidth}
    \centering
    \includegraphics[width=\linewidth]{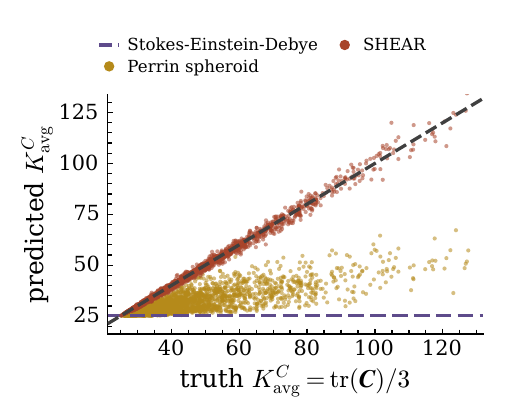}
    \caption{Scalar rotational friction, predicted vs measured.}
    \label{fig:result3_rotation_friction}
  \end{subfigure}\\[4pt]
  \begin{subfigure}[t]{0.31\textwidth}
    \centering
    \includegraphics[width=\linewidth]{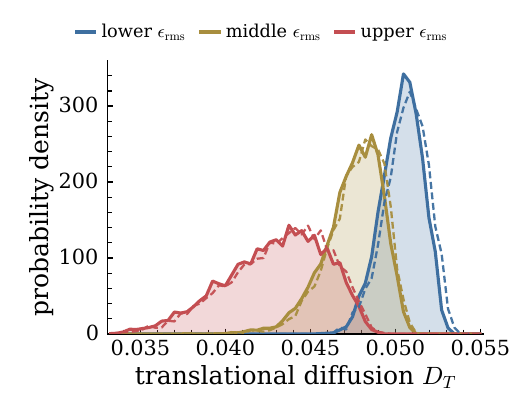}
    \caption{Translational diffusion \(D_T\).}
    \label{fig:result3_brownian_DT}
  \end{subfigure}\hfill
  \begin{subfigure}[t]{0.31\textwidth}
    \centering
    \includegraphics[width=\linewidth]{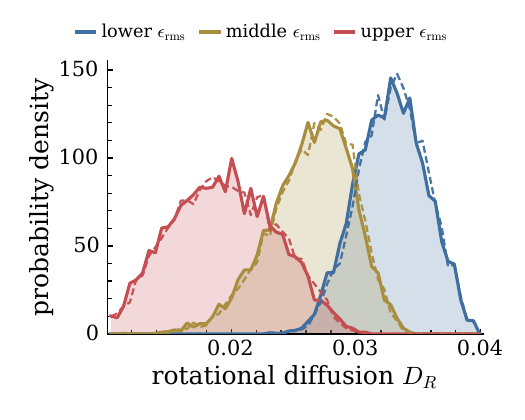}
    \caption{Rotational diffusion \(D_R\).}
    \label{fig:result3_brownian_DR}
  \end{subfigure}\hfill
  \begin{subfigure}[t]{0.31\textwidth}
    \centering
    \includegraphics[width=\linewidth]{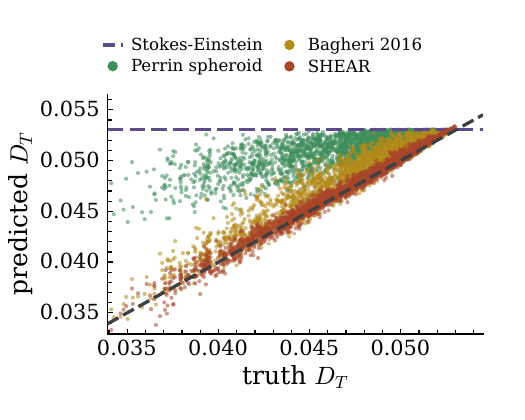}
    \caption{\(D_T\), predicted vs measured.}
    \label{fig:result3_brownian_DT_parity}
  \end{subfigure}
  \caption{\textbf{Orientation-resolved settling, rotation, and Brownian
    dynamics on the \(18\,123\)-shape sealed test set.} Shapes are
    partitioned into equal-population roughness terciles by
    \(\varepsilon_{\rm rms}\), the RMS magnitude of the \(\ell\!\geq\!1\)
    spherical-harmonic log-radius coefficients.
    \emph{Top row --- gravitational settling} (\(64\) orientations per
    shape): (a)~the drift angle between terminal velocity and gravity
    widens with roughness (smooth grains \(\sim\!1.5^{\circ}\), rough
    past \(5^{\circ}\)); (b)~per-shape settling-speed anisotropy
    (\(\sim\!25\,\%\) spread for rough grains); (c)~orientation-averaged
    scalar friction \(K_{\rm avg}\).
    \emph{Middle row --- externally driven rotation} (\(64\) random
    torque directions per shape): (d)~misalignment between angular
    velocity and applied torque (rough grains tail past \(20^{\circ}\));
    (e)~rotational speed anisotropy (rough grains spin two-to-three times
    faster about their easiest axis than their hardest); (f)~scalar
    rotational friction.
    \emph{Bottom row --- Brownian diffusion}: (g)~translational diffusion
    \(D_T\) varies \(\sim\!30\,\%\) across terciles; (h)~rotational
    diffusion \(D_R\) varies by \(\sim\!2.4\times\) (\(R^{-3}\) scaling
    amplifies the shape-induced spread); (i)~\(D_T\) predicted vs
    measured. Throughout, solid curves are the regularised-Stokeslet
    truth and dashed curves the SHEAR surrogate; in the right-hand parity
    panels SHEAR (rust) tracks the diagonal while the classical
    scalar and spheroid reductions collapse to orientation-blind lines
    (Tables~\ref{tab:scalar_friction_A}--\ref{tab:brownian_DT}).}
  \label{fig:result3_grid}
\end{figure*}

The predicted tensors in this study restore exactly this missing
structure. Because
\(\boldsymbol{A}\) is anisotropic, the terminal velocity
\(\boldsymbol{U}=\boldsymbol{A}^{-1}\boldsymbol{F}_{\!g}/\mu\) is in
general not parallel to gravity, so an off-axis grain drifts sideways as
it settles. Across the sealed test set the drift angle has median
\(2.8^{\circ}\) and reaches \(10.7^{\circ}\) for the most irregular
grains (Fig.~\ref{fig:result3_settling_drift}); integrated over the
thousands of kilometres a dust grain is carried, even a few degrees of
persistent lateral bias displaces it well off the vertical fall line
that a scalar parameterisation assumes. The settling \emph{speed} is
orientation-dependent too: over orientations the fastest-to-slowest
ratio has median \(1.17\) and reaches \(1.41\)
(Fig.~\ref{fig:result3_settling_speed}), so a tumbling grain samples a
spread of settling speeds that no single friction coefficient can
express. SHEAR recovers both distributions shape-by-shape.

On the one quantity the correlations \emph{do} target---the
orientation-averaged scalar friction \(K_{\rm avg}\), the comparison
most favourable to them---SHEAR matches the Stokes-flow ground truth to
\(0.6\,\%\) median (\(0.9\,\%\) mean) APE
(Fig.~\ref{fig:result3_settling_friction};
Table~\ref{tab:scalar_friction_A}), and the Stokes-limit reductions of
the correlations do well too: median APE of \(1.7\,\%\)
(Ganser\cite{Ganser1993}), \(2.3\,\%\) (H\"olzer \&
Sommerfeld\cite{Holzer2008}) and \(2.8\,\%\) (Bagheri \&
Bonadonna\cite{Bagheri2016}), with means of \(2.2\)--\(3.2\,\%\). This
result is expected: these are well-calibrated fits to precisely this scalar, so
on it a gross-descriptor correlation and a tensor surrogate stand on
near-equal footing. The difference lies in everything the scalar cannot
carry, which is what the remainder of this section addresses.

\begin{table}[!htbp]
\centering
\caption{Per-shape absolute percentage error (APE) in
  \(K_{\rm avg}=\tfrac{1}{3}\mathrm{tr}(\boldsymbol{A})\)
  on the \(18\,123\)-shape sealed test set: median and mean APE, and
  normalised mean bias.}
\label{tab:scalar_friction_A}
\small
\setlength{\tabcolsep}{4pt}
\begin{tabular}{@{}lccc@{}}
\hline\hline
Method & Median (\%) & Mean (\%) & Bias (\%) \\
\hline
\textbf{SHEAR} & \textbf{0.6} & \textbf{0.9} & \textbf{\(-0.5\)} \\
Ganser 1993 & 1.7 & 2.2 & \(-2.2\) \\
H\"olzer \& Sommerfeld 2008 & 2.3 & 3.1 & \(+3.1\) \\
Bagheri \& Bonadonna 2016 & 2.8 & 3.2 & \(-3.3\) \\
\hline\hline
\end{tabular}
\end{table}

Externally driven rotation shows the same pattern, but more strongly. The
angular velocity misaligns from the applied torque by a median
\(8.1^{\circ}\) (\(P_{90}=21.9^{\circ}\), maximum \(46.0^{\circ}\)), and
the per-shape rotational speed anisotropy has median \(1.48\) and
\(P_{90}=2.50\) (Figs.~\ref{fig:result3_rotation_misalign},
\ref{fig:result3_rotation_anisotropy}). Both exceed their translational
counterparts because the rotational block's cubic size-scaling
magnifies the imprint of shape. On the orientation-averaged rotational friction
(Fig.~\ref{fig:result3_rotation_friction};
Table~\ref{tab:scalar_friction_C}) SHEAR reaches \(1.2\,\%\) median
(\(1.7\,\%\) mean) APE, whereas the isotropic Stokes--Einstein--Debye
sphere \cite{HappelBrenner1983,KimKarrila1991} and the axisymmetric
Perrin spheroid \cite{Perrin1934} under-predict it by \(44.9\,\%\) and
\(31.2\,\%\) (median APE \(36.4\,\%\) and \(23.9\,\%\)). This error is
built into the models rather than a matter of calibration: a triaxial
grain has three distinct rotational eigenvalues, so reducing
\(\boldsymbol{C}\) to a single length (the sphere) or to the two
coefficients of an axisymmetric spheroid (Jeffery\cite{Jeffery1922} and
Bretherton\cite{Bretherton1962}) forces at least two of the three to
coincide and cannot represent the third.

\begin{table}[!htbp]
\centering
\caption{Per-shape absolute percentage error in
  \(K_{\rm avg}^{C}=\tfrac{1}{3}\mathrm{tr}(\boldsymbol{C})\): median
  and mean APE, and normalised mean bias.}
\label{tab:scalar_friction_C}
\small
\setlength{\tabcolsep}{4pt}
\begin{tabular}{@{}lccc@{}}
\hline\hline
Method & Median (\%) & Mean (\%) & Bias (\%) \\
\hline
\textbf{SHEAR} & \textbf{1.2} & \textbf{1.7} & \textbf{\(-0.6\)} \\
Perrin spheroid & 23.9 & 26.3 & \(-31.2\) \\
Stokes--Einstein--Debye sphere & 36.4 & 38.1 & \(-44.9\) \\
\hline\hline
\end{tabular}
\end{table}

Brownian diffusion completes the picture. Across the roughness terciles
(Figs.~\ref{fig:result3_brownian_DT}--\ref{fig:result3_brownian_DT_parity};
Table~\ref{tab:brownian_DT}), \(D_T\) varies by \(\sim\!30\,\%\)
and \(D_R\) by a factor of \(\sim\!2.4\). SHEAR reduces the
per-shape error on \(D_T\) from \(2.4\)--\(11.1\,\%\) median APE
(\(2.9\)--\(13.8\,\%\) mean; the range spanned by the empirical and
spheroid baselines) to \(0.6\,\%\) median (\(0.9\,\%\) mean).
Across settling, rotation, and Brownian motion alike, the tensor
surrogate tracks the regularised-Stokeslet solver to within about one
percent, while the sphere and spheroid baselines err by tens of percent
on exactly the anisotropic, orientation-dependent quantities that
dominate each process.

\begin{table}[!htbp]
\centering
\caption{Per-shape absolute percentage error in
  \(D_T=(k_B T/3\mu)\,\mathrm{tr}(\boldsymbol{A}^{-1})\): median and
  mean APE, and normalised mean bias.}
\label{tab:brownian_DT}
\small
\setlength{\tabcolsep}{4pt}
\begin{tabular}{@{}lccc@{}}
\hline\hline
Method & Median (\%) & Mean (\%) & Bias (\%) \\
\hline
\textbf{SHEAR} & \textbf{0.6} & \textbf{0.9} & \textbf{\(+0.5\)} \\
Bagheri \& Bonadonna 2016 & 2.4 & 2.9 & \(+2.7\) \\
Perrin spheroid & 7.6 & 9.6 & \(+9.1\) \\
Stokes--Einstein sphere & 11.1 & 13.8 & \(+13.1\) \\
\hline\hline
\end{tabular}
\end{table}

\section{Discussion}\label{sec:discussion}

\paragraph{Spectral fidelity: scope and outlook.}
Result~1 reduces the question ``is an \(\ell\leq 15\)
spherical-harmonic representation enough?'' to numbers: the median
cost of the representation is \(0.14\%\) (translation) and \(0.38\%\)
(rotation) of the resistance, \(94\%\) and \(81\%\) of random
particles are inside the \(5\%\) engineering tolerance, and the
exceptions concentrate toward the rough, shallow-decay corner of the
\((\alpha,\,\varepsilon_{\rm rms})\) plane, where the phase maps of
Fig.~\ref{fig:result1_main}c,d supply the required higher truncation
at a glance. The underlying physics --- the truncation error follows
the kernel-weighted discarded power
(eqn~\eqref{eq:kernel_tail}), with per-mode sensitivities that grow
with \(\ell\) and a measured \(K_C/K_A\) asymmetry approaching \(3\) that makes
rotation the binding block --- is established and validated in the
ESI (Section~S1). The same analysis shows where every
power-spectrum-only summary must stop: on anisotropic bases the
roughness acquires an \emph{orientation-dependent} linear response
that \(P_\ell\) cannot encode, and which is recoverable only from the
full, orientation-resolved coefficients. That observation is the
conceptual bridge to Result~2: the equivariant surrogate consumes
exactly the information the spectral summary discards.

Three caveats bound the scope. First, the kernel law is a leading-order
result: each roughness mode enters at second order in its amplitude (the
linear term vanishes by the selection rule) and the modes are summed
independently. For shapes with \(\varepsilon_{\rm rms}\gtrsim 20\%\) the
neglected mode-coupling and higher-order terms grow and the
proportionality in eqn~\eqref{eq:kernel_tail} drifts, so the law
misestimates \emph{how many} modes a very rough shape needs. It still
ranks shapes correctly by truncation error, however, so a shape that
needs more modes than another is still identified as such. Second, the
framework assumes star-convex surfaces, the natural domain of a radial
spherical-harmonic expansion. Third, the single exponent~\(\alpha\)
fitted on \(\ell\in[3,20]\) is a coarse summary of the spectral shape,
and bimodal or knee-shaped spectra are not faithfully captured by it.

\paragraph{Surrogate accuracy: limits and outlook.}
SHEAR predicts the translational and rotational resistance blocks to
within \(1.4\,\%\) and \(2.7\,\%\) with fewer than \(10^{6}\) parameters,
while remaining exactly equivariant to machine precision. The residual
error is concentrated in the high-roughness class
(\(D_{9{-}15}\geq 0.10\)), whose rotational block
reaches \(3.7\,\%\). The low-roughness band
(\(D_{9{-}15}<0.05\), which includes the smooth ellipsoids)
already sits below \(1.8\,\%\). This residual tracks the surface
roughness carried by spherical-harmonic degrees \(\ell\geq 9\) (the
\(D_{9{-}15}\) band), which lie \emph{above} SHEAR's effective input
band-limit of \(\ell\approx\ell^{\mathrm{tp}}_{\max}=8\)
(Section~\ref{sec:model}): the network cannot discern that roughness
directly, so on rough grains it must infer the \(\ell\geq 9\)
contribution to the resistance, and the irreducible part of that
inference sets the floor on its accuracy---closely related to the
degree-8 truncation error of Result~1. It is therefore in part a
property of the input representation and the resolution-4 labels rather
than of the trunk. Extending the equivariant tensor-product input to
higher \(\ell\) is the natural route to reducing it, though the
ablation of Section~S5 (raising \(\ell^{\mathrm{tp}}_{\max}\) from
\(8\) to \(10\)) shows the gain over such a modest extension is already
within run-to-run noise. Capturing the full \(\ell\leq 15\) tail would
require both a higher ceiling and finer solver labels. The purely equivariant trunk is, moreover, the
right design choice rather than a sacrifice for compactness: it
outperforms substantially larger, invariant-augmented variants
(Section~S5), because the rotation-invariant content those hand-built
features encode is already reachable inside the trunk, and supplying it
redundantly only enlarges the optimisation problem. The comparison with the invariant baseline
sharpens where equivariance matters: not in the eigenvalue magnitudes,
which SHEAR predicts at least as accurately as a model built solely from
invariants, but in the orientation of the tensors, which is inaccessible
to any rotation-invariant method and is exactly what the downstream
transport observables of Section~\ref{sec:settling_results} require.
Finally, although SHEAR already carries a mixed-parity hidden basis, its
output heads are parity-even and it predicts only \(\boldsymbol{A}\) and
\(\boldsymbol{C}\); adding a parity-odd (\(1{\times}2^{o}\)) output head to
predict the chirality-sensitive coupling block \(\boldsymbol{B}\)---which
the mixed-parity trunk is already equipped to feed---together with
broadening the training distribution and a systematic characterisation of
out-of-distribution generalisation, are the natural next steps.

\paragraph{Settling and rotation: limits and outlook.}
The settling analysis uses the frozen-orientation approximation, which
is exact within the parity-even (\(\boldsymbol{B}=0\)) scope of this
study. For particles with a non-trivial coupling block
\(\boldsymbol{B}\), the deterministic trajectory becomes helical. The
parity-odd extension of the surrogate to predict \(\boldsymbol{B}\)
is the natural next step. Coupling the body to a background shear
flow requires the additional strain-coupling block of the
\(11\times 11\) Faxén-extended grand mobility, which the \(6\times 6\) GRM
does not predict. Modelling that channel is future work.

The rotational-friction comparison is limited to two classical
baselines (sphere and Perrin spheroid) because no empirical
correlation for orientation-resolved rotational friction of irregular
particles currently exists in the literature — a gap that the present
surrogate is designed to fill.

\section{Conclusions}\label{sec:conclusions}

The goal of this study is to replace the scalar drag coefficient---the
single number to which decades of practice have compressed the
hydrodynamics of an irregular microparticle---with the object the
physics actually requires: the full, anisotropic translational and
rotational resistance tensors. Three results, taken together, show
that this is both possible and practical.

The first result establishes how faithfully a particle's shape must be
resolved before its resistance is fixed. Truncating a shape's
spherical-harmonic expansion at degree~\(L\) leaves an error in the
grand resistance matrix that is set, on both the translational and
rotational blocks, by the \emph{kernel-weighted} discarded power: each
lost harmonic mode contributes in proportion to its own per-mode Stokes
sensitivity \(K_X(\ell)\). Measured directly from single-mode sphere
perturbations, the rotational-to-translational ratio \(K_C/K_A\)
approaches~\(3\), so the rotation block always
requires more harmonics than the translation block---two to three times
as many at a fixed tolerance. This converts an open practical question
(how many modes does a given shape need?) into a single lookup over
the shape's roughness and spectral slope, and is, to our knowledge, the
first empirical spectral-convergence law for the Stokes resistance
matrix.

The second result is SHEAR, an \(\mathrm{SO}(3)\)-equivariant neural
surrogate that maps a particle's spherical-harmonic shape directly to its
symmetric positive-definite \(\boldsymbol{A}\) and \(\boldsymbol{C}\)
blocks. Because every layer is an equivariant primitive, a rotated
particle yields the exactly-rotated tensor---to floating-point
precision, with no data augmentation and no symmetry left to be learned
approximately. With fewer than \(10^{6}\) parameters SHEAR predicts the
two blocks to \(1.4\,\%\) and \(2.7\,\%\) mean relative error on a
sealed test set of \(18\,123\) shapes, in tens of microseconds per
shape---four to five orders of magnitude faster than the
regularised-Stokeslet-surface solver that generated its labels. To our knowledge it is the first
equivariant neural surrogate for the complete anisotropic resistance
tensors of arbitrarily shaped microparticles, and it is accurate
exactly where scalar correlations and rotation-invariant networks
cannot be: in the orientation of the tensors, not merely their
magnitude.

Finally, the third result shows why that orientation matters. Across over a
million orientation-sampled settling, rotation, and diffusion events,
the surrogate recovers lateral settling drift of up to~\(11^{\circ}\),
rotational misalignment with the applied torque up to~\(46^{\circ}\), and
shape-induced spreads of \(30\,\%\) in translational and a factor of
\(\sim\!2.4\) in rotational diffusion---every one of which is identically
zero under a scalar or spheroid reduction. On the orientation-averaged
friction --- the single quantity the classical correlations target ---
their Stokes-limit reductions are themselves accurate
(\(1.7\)--\(2.8\,\%\) median APE per shape) and SHEAR is comparable
(\(\sim\!1\,\%\)). These are excellent finite-Reynolds drag fits. The
decisive point is that they are orientation-blind scalars by
construction, so the drift, misalignment, and diffusion anisotropy above
are inaccessible to them entirely. A fast tensor surrogate does not
merely sharpen the scalar answer. It supplies physics that the scalar
answer structurally cannot.

Several extensions follow naturally. The most immediate is chirality:
SHEAR already carries a mixed-parity hidden basis, so a single
parity-odd output head would let it predict the coupling block
\(\boldsymbol{B}\) and, with it, the helical settling and
handedness-dependent separation of genuinely chiral particles that the
present parity-even scope deliberately sets aside. A second direction
is ambient flow: promoting the target from the \(6\times6\) resistance
matrix to the \(11\times11\) Fax\'en-level grand mobility would capture the
strain coupling a particle experiences in shear, opening the surrogate to
suspension rheology and to the Jeffery-orbit dynamics of non-spherical
grains. Finally, the single-particle tensors learned here are the exact
building block that many-body and continuum models presuppose: pairing
SHEAR with a graph-network many-body correction, and validating it
against \(\mu\)CT-scanned mineral dust and volcanic ash, would carry
orientation-resolved hydrodynamics from individual grains to the
population-scale transport problems---atmospheric dust, microplastic
fate, and colloidal assembly---that motivated this work.

The deepest opportunity, however, is dynamical. A boundary-element
solve fixes the mobility of \emph{one} frozen shape in seconds to
minutes. The most consequential microhydrodynamic processes are
precisely those in which the shape itself \emph{evolves}, and following
that evolution has been computationally out of reach. Mineral grains
and microplastics round and roughen as they are abraded, weathered, and
dissolved, tracing a particle trajectory through morphology space whose drag and
settling history controls sediment provenance, atmospheric residence
time, and the depth to which a fragment is carried. Flexible fibres,
filaments, polymer chains, and biological aggregates continuously bend
and reconfigure, so that every conformation along a deformation
trajectory is a new shape demanding a fresh mobility tensor---the
central bottleneck in modelling fibre-suspension rheology and the
transport of elongated microplastics. And aggregation runs the process
in reverse: soot, snow, marine snow, flocs, and colloidal and protein
assemblies grow by particles joining into ever-larger composites whose
hydrodynamic response must be recomputed as they assemble and fragment.
In all of these, a microsecond-per-shape, fully anisotropic surrogate
turns the mobility tensor from a one-off precomputation into an
inexpensive function call inside the dynamics---a ``mobility in the
loop'' that lets shape evolution and hydrodynamic transport be advanced
together over the \(10^{6}\)--\(10^{9}\) shape-timestep evaluations such
simulations require. By making the cost of a shape-resolved mobility
negligible, equivariant surrogates open low-Reynolds-number
hydrodynamics to the entire class of problems in which morphology is not
a fixed input but a dynamical variable.

\section*{Author contributions}
S.~Pradeep: conceptualization, methodology, software, validation, formal
analysis, investigation, data curation, visualization, and writing --
original draft. D.~Dandy: writing -- review and editing.
C.~S.~J.~Tsai: funding acquisition and writing -- review and editing.
J.~D.~Eldredge: conceptualization, supervision, resources,
and writing -- review and editing.

\section*{Conflicts of interest}
There are no conflicts to declare.

\section*{Data availability}
The shape-generation, regularised-Stokeslet solver, and equivariant-surrogate
(SHEAR) source code, together with the scripts that regenerate every figure in
this article, are available at
\url{https://github.com/SanjayPradeep97/Stokes-Hydrodynamics-of-Arbitrary-Random-Particles}.
The generated dataset of spherical-harmonic shape coefficients with their
grand-resistance-matrix labels, and the trained surrogate weights, are archived
at [DOI to be assigned upon acceptance].

\section*{Acknowledgements}
This work used computational and storage services associated with the Hoffman2 Shared Cluster provided by
UCLA Office of Advanced Research Computing's Research Technology Group.

\bibliographystyle{rsc}
\bibliography{references}

\end{document}